\documentclass[fleqn,usenatbib,twocolumn]{mnras}

\usepackage{amssymb}
\usepackage{amsmath}
\usepackage{graphicx}
\usepackage{xcolor,ulem}
\usepackage{multirow}

\title[Contemporaneous GRB afterglow]{The contemporaneous phase of GRB Afterglows --- Application to GRB~221009A}

\author[E. Derishev and T. Piran]{Evgeny  Derishev$^{1}$ and Tsvi Piran$^{2}$ 
\\
$^{1}$Lobachevsky State University of Nizhny Novgorod, 23 Gagarin av, 603950 Nizhny Novgorod, Russia\\
$^{2}$Racah Institute of Physics, The Hebrew University of Jerusalem, Jerusalem 91904, Israel }

\date{Accepted XXX. Received YYY; in original form ZZZ}

\pubyear{2023}

\begin{document}

\newcommand{\diff}[1]{\mathop{}\!\mathrm{d} #1}

\label{firstpage}
\pagerange{\pageref{firstpage}--\pageref{lastpage}}
\maketitle

\begin{abstract}
The TeV observations of GRB~221009A provided us with a unique opportunity to analyze the contemporaneous phase  in which both  prompt and afterglow emissions are seen simultaneously. 
To describe this initial phase of Gamma-Ray Burst afterglows, we suggest a model for a  blast wave with intermittent energy supply.
We treat the blast wave as a two-element structure. 
The central engine supplies energy to the inner part (shocked ejecta material) via the reverse shock.  
As the {shocked ejecta material} expands, its internal energy is transferred to the shocked external {matter}.
We take into account the inertia of the shocked external material so that the pressure difference across this region determines the derivative of the blast wave's Lorentz factor.
Applied to GRB~221009A, the model  yields a very good fit to   the observations  of  the entire TeV lightcurve except for three regions where there are excesses in the data with respect to the model. Those are well correlated with the three largest episodes of the prompt activity
and thus, we interpret them as the reverse shock emission.
Our best-fit solution for GRB~221009A is an extremely narrow jet with an opening angle {$\theta_j \approx 0.07^{\circ} (500/\Gamma_0)$} propagating into a wind-like external medium.
This extremely narrow angle is consistent with the huge isotropic equivalent energy of this burst  and its inverse jet break explains the  very rapid rise of the afterglow. Interestingly, photon-photon annihilation doesn't play a decisive role in the best-fit model. 
\end{abstract}


\section{Introduction} 
\label{sec:intro}

Gamma Ray Bursts (GRBs), see e.g. \citep[GRBs, see e.g.][for a review]{Piran2004}  are observed across the entire electromagnetic spectrum --- from radio frequencies to TeV energies. Nevertheless, many important aspects of their physics remain poorly understood and every out-of-the-ordinary observation offers new insights.

GRB emission is divided into two phases: prompt and afterglow. The prompt emission is often rapidly variable with a complex lightcurve, whose structure is attributed to the activity of the central engine.
It is followed by a gradually declining afterglow emission with a smooth lightcurve, which is attributed to a relativistic blast wave that forms when the material ejected by the central engine during the prompt phase interacts with  the circumburst medium. It is natural to expect that the early  afterglow  
(i.e., the emission from the external shock) overlaps in time with the prompt emission \cite[see e.g.][]{Zou2010}. However, this overlapping initial stage of afterglows is hidden from view: afterglows' X-ray signal is indiscernible against a much stronger prompt X-ray signal.

High-energy gamma-ray emission from GRBs was predicted before actual detections. From efficiency arguments, it was anticipated that GRBs have synchrotron-self-Compton (SSC) component in the TeV range and that it carries about 10 percent of the emitted energy \citep{DKK2000,DKK2001}. Early attempts to estimate the parameters of the high-energy SSC component from the first principles placed it in the GeV band \citep{SariEsin2001}.   
Numerous other efforts to estimate this emission 
\citep{ZhangMes2001B,Razzaque2004,Pe'er2005,Fan2006,Fan2008,Panaitescu2008} followed.
Later the expectations were extended to the TeV band 
{for dense stellar wind environments}
\citep{VurmBeloborodov2017}. It was not until the appearance of the pair-balance model \citep{DerishevPiran16} that first-principles predictions of the strength of the SSC component became possible. The pair-balance model places the peak of the SSC component at around {1~} TeV 
(rather insensitive to GRB parameters) and predicts that its power is always comparable to the power of the synchrotron component.

When Fermi-LAT detected early GeV emission from several GRBs,  it was suggested \citep{Kumar2010,Ghisellini2010}  that this is an afterglow and not prompt emission, even though there was some overlap of the observed GeV signal with the lower energy gamma-rays. However, there was not enough data to explore the contemporaneous prompt-afterglow  phase.
Recently, the TeV observatory  LHAASO \citep{LHAASO} detected a strong TeV  signal while numerous satellites \citep{KonusWind,HXMT,FermiGBM} 
measured a simultaneous  low-energy gamma-rays from   GRB 221009A. The 
lightcurve in low-energy gamma-rays shows the usual complex structure with several pulses and rapid variability, whereas the TeV lightcurve is smooth as expected for the signal from an expanding blast wave.  It should be noted that GRB~221009A was the brightest so far, and there is a good chance that its signal-to-noise ratio in the very first unambiguous afterglow onset detection will remain unbeatable for a long time.

{The unique observations of GRB~221009A have already attracted a lot of attention, and several authors attempted to explain different features of this burst \citep{Zhang_etal2023,OConnor2023,Khangulyan2023,Ren2023,Sato2023}.} 
Here we focus on the fact that 
the observation of the initial afterglow stage, overlapping with the prompt phase, offers a unique opportunity to extract information that would be inaccessible by other means. In particular, this includes the density profile of the circumburst medium as well as some characteristics of the central engine, such as the width of the jet, the variability of its Lorentz factor, and the radiative efficiency during the prompt phase.

A model for the earliest afterglow has to take into account that the blast wave continuously acquires more energy from the central engine. 
So far, attention has been given to the behaviour of the reverse shock in the presence of the variable energy supply from the ejecta 
\cite{Nakar2004,Nakar2005,McMahon2006}.  This is natural as the reverse shock is the one that directly interacts with the ejecta. Less attention was given to the impact  of variable energy supply on the forward shock. 
Indeed, models of blast waves with energy supply (and leakage due to radiation) have been known for a long time: the seminal \cite{BlandfordMcKee} paper describing a   self-similar solution {(hereafter the BM solution)} of a constant energy blast wave  also presented blast wave solutions with increasing energy. 
Various authors considered plain archetypal scenarios: continuous energy injection at a rate that varies as a power of time \citep{CohenPiran1999,ZhangMes2001A,SariK2000} or impulsive energy injection \citep{SariK2000,KumarPiran2000}. Other works simply considered a BM solution in which the energy increases as a function of time. 
However, these models, which assume that a  self similar solution has already been established, are not suitable for the very early phase in which  the reverse shock  is still active and it is gradually transferring its energy to the forward shock.  
The case of GRB~221009A,  with  simultaneous measurements of both the central engine's activity and the emission from the blast wave fed by this activity, enables us to explore this stage.  A suitable theory is necessary to use these observations and build a coherent physical concept of the earliest afterglow.

Our goal is to construct a minimalistic (i.e., as simple as possible) but functional (i.e., verified against observations) model for an expanding relativistic blast wave that is continuously fed with energy from the central engine at an arbitrary time-dependent rate.
Observation of the earliest afterglow from GRB~221009A provides a testbed for such a model. Incorporating model elements one by one and comparing the theoretical results to the observations, we were able to justify that our model is indeed the minimal functional model.

Our model is based on the common reverse shock, forward shock and contact discontinuity structure. Namely, at the initial phase, the blast wave is composed of a shocked ejecta material and a shocked external material that are in pressure balance at the contact discontinuity between them. We include two new ingredients to this picture. First, noticing that the energy from the ejecta is supplied through the reverse shock to the shocked ejecta material which gradually pushes on the shocked external material, giving the forward shock additional energy,  we consider the energy transfer between those two regions. Second, we explicitly take into account the inertia of the shocked external matter, by relating the  blast wave's acceleration (or  deceleration) to the pressure gradient inside. 
We treat both regions as single elements and therefore ignore all possible internal dynamics.
Although we do not resolve the spatial structure of the forward shock region, the pressure difference across it is taken into account, being an essential constituent of the model.
To reach a better quantitative agreement with observations at later times one needs to consider a narrow jet instead of a spherical outflow, or otherwise the luminosity decline is too slow.

This model provides a good fit to the measured TeV lightcurve of GRB~221009A, which likely makes it a valid choice for modeling the afterglow onset phase in other GRBs as well. In addition, the model serendipitously produced a fairly convincing case for emission coming from the reverse shock. The reverse-shock emission component turns out to be (i) much weaker than emission from the forward shock and (ii) in principle discernible in TeV lightcurves.
Although the reverse shock contribution appears to be small in the TeV range, it may be stronger in other spectral bands \citep{Sari1999RS,Sari1999RSB,Meszaros1999,Wang2001,Nakar2004,Razzaque2004,Genet2007}.

The paper is organized as follows.
In Sect.~(\ref{sec:model}) we present the new model for relativistic blast wave with continuous {and intermittent} energy supply, and discuss solutions for afterglow lightcurves for several typical scenarios. In Sect.~(\ref{sec:lightcurve_fit}) we 
describe the method of lightcurve modeling. We then apply the model to TeV observations of GRB~221009A. We find an excellent best-fit solution and we analyze its properties.
In Sect.~(\ref{sec:Discussion}) we list our most important findings also mentioning possible alternative interpretations.

\section{GRB~221009A observations} 
\label{sec:observations}

\begin{figure}
    \centering
    \includegraphics[width=1.0\linewidth]{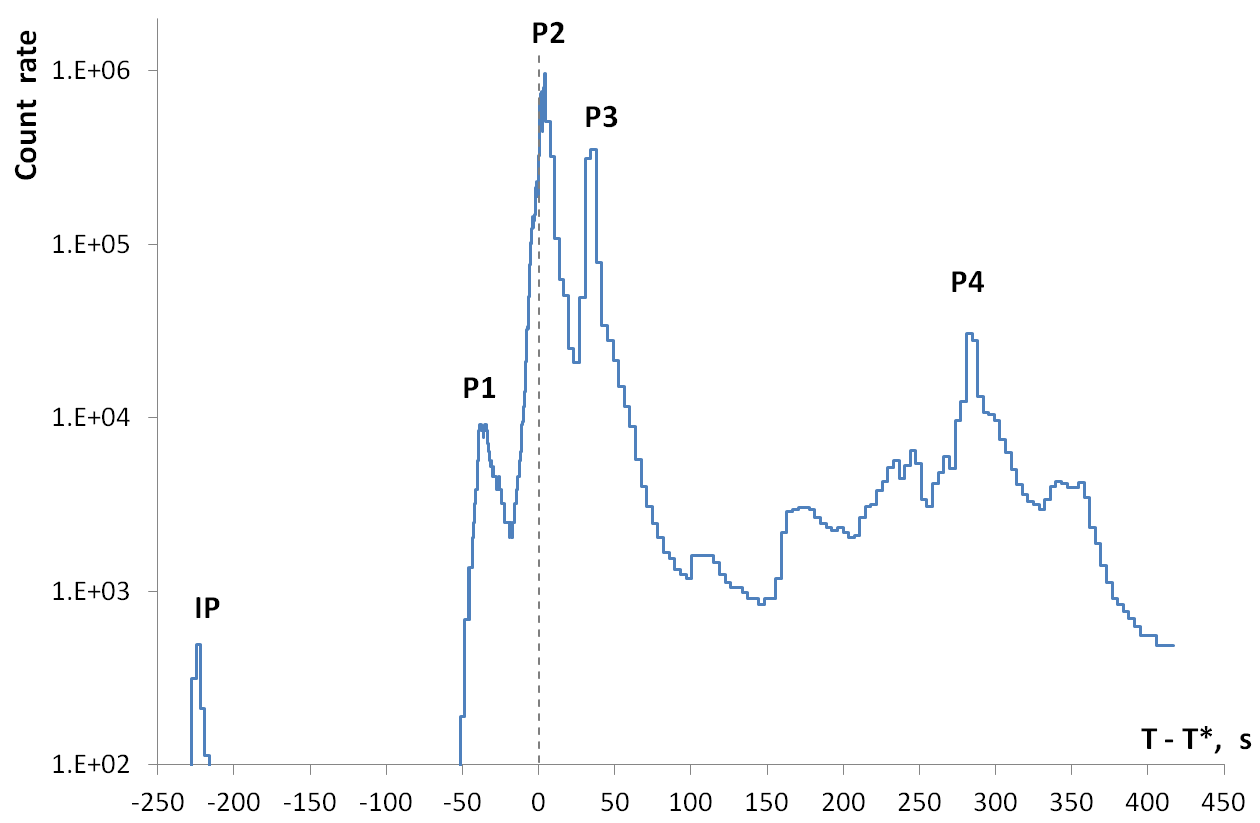}
    \caption{
    The count rate for the prompt phase of GRB~221009A, compiled from Konus-WIND measurements \citep{KonusWind}. 
    Our assumption is that the prompt kinetic luminosity follows the count rate.
    }
    \label{fig:Xray_lightcurve}
\end{figure}

The prompt emission of GRB~221009A was so bright, that it saturated all the X-ray telescopes that attempted to detect it. However, Konus-WIND and ART-XC teams \citep{KonusWind}  were able to reconstruct the prompt lightcurve.
combining their measurements. 
Figure~(\ref{fig:Xray_lightcurve}), which is compiled from the \cite{KonusWind} data, presents the prompt lightcurve. A similar lightcurve was obtained by and GECAM-C
\citep{HXMT}.
The three brightest pulses together contain about 99 percent of the entire energy release during the prompt phase of GRB~221009A. Their energy shares are in proportion 6:3:1. We will name these pulses  P2, P3, and P4, following the notation of \cite{KonusWind}.

\begin{figure}
    \centering
    \includegraphics[width=1.0\linewidth]{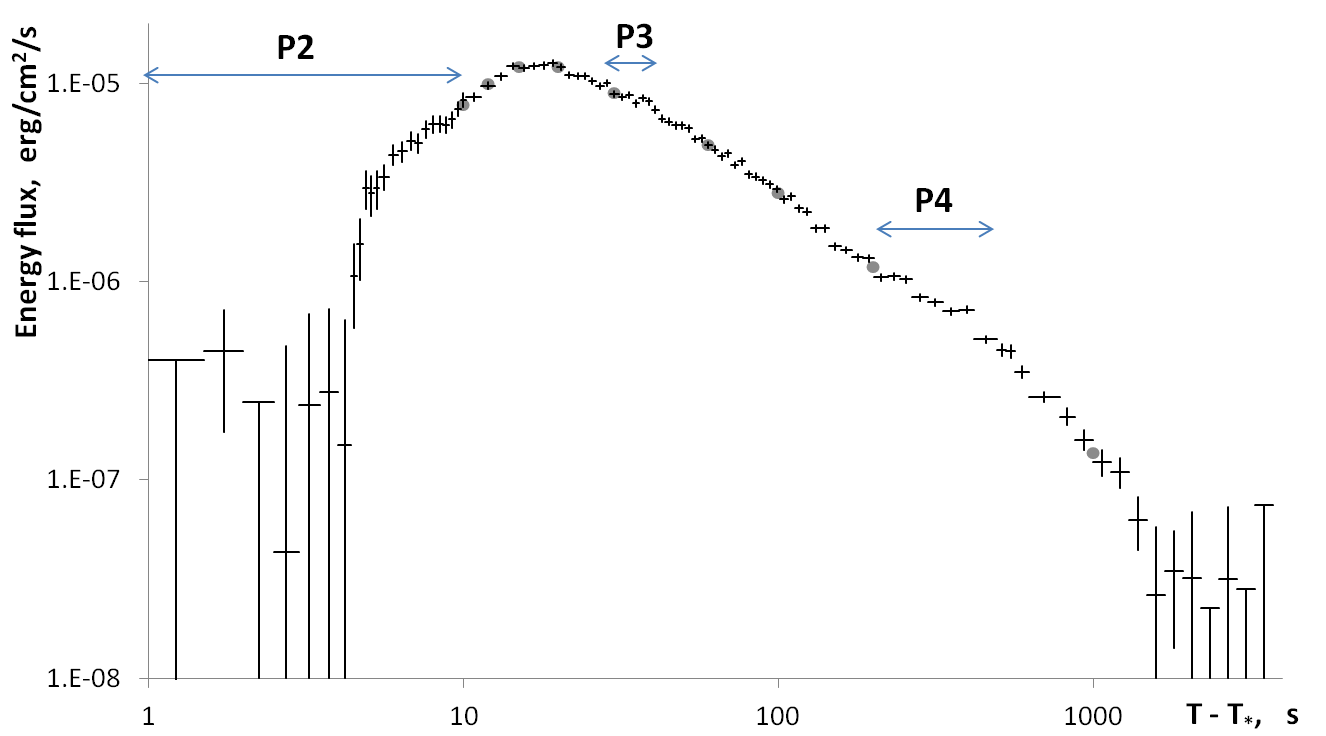}
    \includegraphics[width=1.0\linewidth]{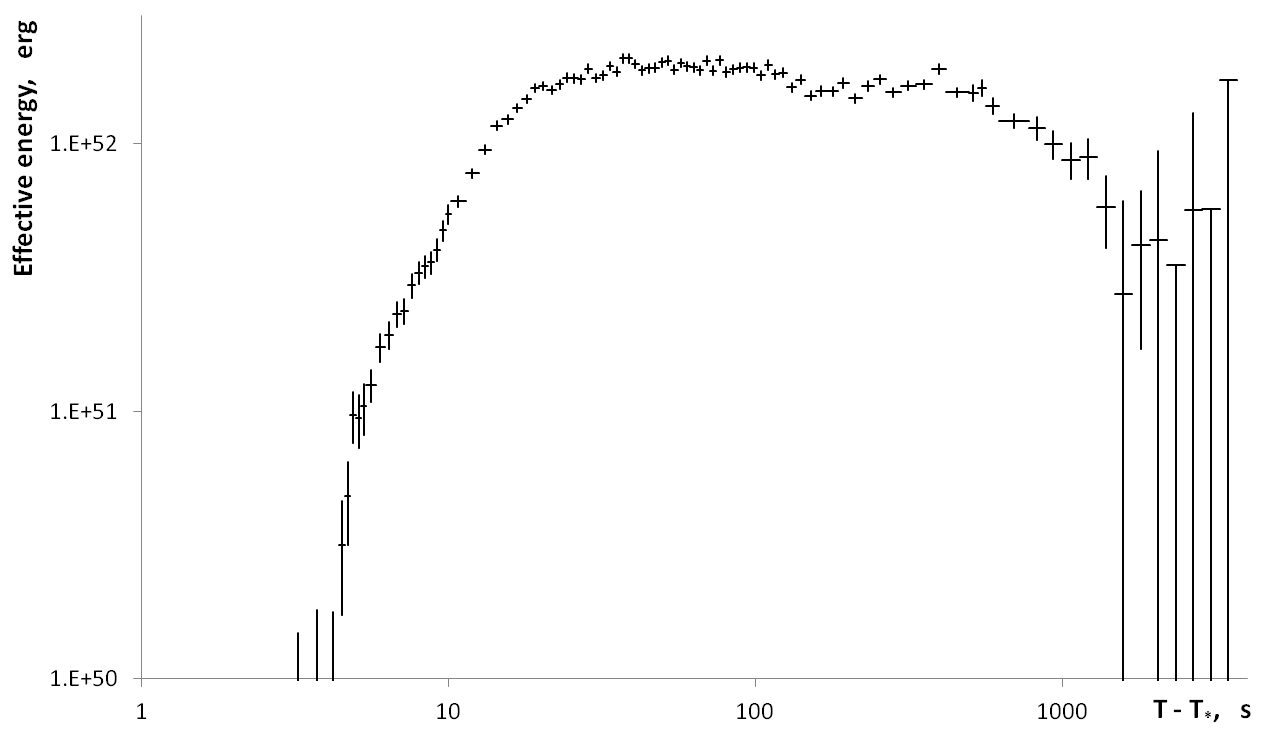}
    \caption{
    Upper panel: the lightcurve of GRB~221009A (energy flux $F$ integrated over the energy range from 0.3 to 5~TeV).
    Lower panel: the effective blast wave energy, $E_\mathrm{eff} = 4\pi D_{L}^2 F \times (T - T_*)$.
    The gray circles indicate reference points adopted to estimate quality of lightcurve fits (see Sect.~\ref{sec:lightcurve_fit}). 
    These points are located at 10~s, 12~s, 15~s, 20~s, 30~s, 60~s, 100~s, 200~s, and 1000~s.
    }
    \label{fig:TeV_lightcurve}
\end{figure}

TeV lightcurve of GRB~221009A was reported by \cite{LHAASO} and is shown in the upper panel of Fig.~(\ref{fig:TeV_lightcurve}). To make the lightcurve less noisy we combined the original data points between 10~s and 500~s into sets of three, while the other data points remain unchanged. Following \cite{LHAASO}, we measure time from $T_* \equiv T_0 + 226$~s, where the TeV signal appears first, with the sharp rise that is delayed by several seconds with respect to the beginning of the strongest pulse in the prompt lightcurve. Unlike the prompt low-energy gamma-ray emission, the TeV lightcurve is smooth and almost featureless. 
It has a rather sharp peak at approx. 17~s and two barely distinguishable features (local excesses) in its decaying branch, one narrow feature at approx. 40~s and one broad feature between approx. 250~s and 600~s.

The drastic difference between the rapidly and strongly variable prompt {low-energy gamma-rays}  and the smooth TeV lightcurve (the two do not show any apparent correlation)  immediately led to the conclusion that, despite the temporal coincidence with the prompt {low-energy} gamma-rays, the TeV emission is due to afterglow and it arises from the  blast wave expanding into the circumburst medium \citep{LHAASO}.

For an afterglow, it is convenient to complement the lightcurve (i.e., temporal dependence of energy flux $F(T)$) with another plot for quantity $E_\mathrm{eff} = 4\pi D_{L}^2 F \times (T - T_*)$, that is source luminosity multiplied by observer's time. For a self-similar blast wave, the bolometric luminosity is $L_\mathrm{bol} = (1+z) \epsilon_{r} C_\mathrm{_L} E^\mathrm{(iso)}_\mathrm{kin}/t_\mathrm{obs}$ 
\citep{CohenPiran1999,Derishev23}, where $\epsilon_{r}$ is radiative efficiency, $E^\mathrm{(iso)}_\mathrm{kin}$ the shock's kinetic energy (isotropic equivalent), and the numerical coefficient $C_\mathrm{_L}$ is of the order of unity (it depends on the external density profile). Therefore, the quantity $E_\mathrm{eff}$ has the meaning of the proxy to the blast wave energy (we will call this effective blast wave energy). 
It is plotted in the lower panel of Fig.~(\ref{fig:TeV_lightcurve}). 
The effective energy reaches its very broad maximum at approx. 70~s, i.e. much later than the lightcurve's maximum. Apparently, this fact reflects the history of energy injection into the blast wave. 
A subsequent decline in $E_\mathrm{eff}$ may be either due to the deceleration of the  structured-jet blast wave or due to radiative energy losses, or both. The former brings into view parts of the jetted structure that move at larger angles to the line of sight and have lower isotropic equivalent 
energy, thus reducing the average isotropic equivalent energy. 
The two features (excesses) in the lightcurve,  described earlier, become more visible in the effective energy plot.

\section{ A relativistic blast wave model with an intermittent energy supply} 
\label{sec:model}

\subsection{The model}
\label{sec:model_construction}

Put in simple words, the observed lightcurve of GRB~221009A afterglow (see Fig.~\ref{fig:TeV_lightcurve}) poses two challenges for any blast wave model that attempts to describe it. 
First, the model has to explain the very rapid rise of the lightcurve, where the rise timescale $t_\mathrm{rise}$ is much smaller than the decay timescale $t_\mathrm{decay}$.
Second, the lighcurve has a very simple shape, with its maximum located between two dominant episodes of prompt emission (these must be as well the main episodes of energy injection to the blast wave), showing no visible response to the second episode. With some caution, one can interpret this combination of features in the following way: the blast wave is much more responsive to energy input when it accelerates than when it decelerates.

In this section we describe the minimalistic blast wave model that can reproduce observations. 
Consider, first, the classical BM solution. The energy of the blast wave is: 
\begin{equation} \label{piston_model}
    E \equiv \Gamma_0 M_\mathrm{ej} c^2 
    =  C_\mathrm{_E} \Gamma_\mathrm{sh}^2 M_\mathrm{fs} c^2 \, ,
\end{equation}
where $M_\mathrm{ej}$ is the {mass of the shocked ejecta}, $M_\mathrm{fs}$ the swept-up mass, $\Gamma_0$ the Lorentz factor of ejected material at the moment of ejection, and
the factor $C\mathrm{_E} $ is a numerical factor that depends on the density profile of the external medium. 
In the case of ISM (a constant density) circumburst medium $C_\mathrm{_E}= 6/17$, and in the case of wind-like ($\rho \propto R^{-2}$) circumburst medium $C_\mathrm{_E}= 2/9$. Radiative losses change the blast wave's deceleration law, thus altering the coefficient $C_\mathrm{_E}$ \citep{PartiallyRadiativeShock}. 
We restrict our analysis to power-law density profiles ($\rho \propto R^{-k}$) that allow us to relate the swept mass and the local density
\begin{equation}  \label{rho_Msw_relation}
    M_\mathrm{fs} = \frac{4\pi}{3-k} R^3 \rho  \, .
\end{equation}

It is straightforward to take into account  energy injection by the kinetic energy of the ejecta and radiative losses \citep[generalizing][]{CohenPiran1999}:
\begin{equation} \label{energy_input}
    E = \int_0^{t_\mathrm{inj}} L_\mathrm{kin} \diff{t} - 2\pi \int_0^{R} \epsilon_\mathrm{r} \Gamma_\mathrm{sh}^2 \rho c^2 \left( R^{\prime} \right)^2 \diff{R^{\prime}}      \, .
\end{equation}
Here 
$\epsilon_\mathrm{r}$ is the fraction of shocked gas energy that is converted into radiation and  
\begin{equation} \label{injection_time}
    t_\mathrm{inj} (R) = \int_0^{R} \frac{\diff{R^{\prime}}}{2\Gamma_\mathrm{sh}^2 c}  - \frac{R}{2\Gamma_0^2 c}      \, .
\end{equation}
is the difference between blast wave propagation time and free coasting time for the ejected material, calculated in ultrarelativistic approximation.
Note that if $\Gamma_0$ depends on $t_\mathrm{inj}$, then Eq.~(\ref{injection_time}) may have more than one solution. In this case, internal shocks form within the ejected material at the coasting phase. We do not consider this possibility.

The model described by Eqs.~(\ref{piston_model}-\ref{injection_time})
fails to explain the afterglow of GRB~221009A --- it proves to be too responsive to energy injection and it always produces a prominent bump following the second episode of prompt activity. The reason for such behavior is the absence of inertia: whenever energy is injected into the blast wave, the latter immediately reacts, increasing its Lorentz factor.

\cite{DermerHumi2001}, revisited by \cite{NavaSironi2013},  
suggested 
that the rate at which the blast wave's internal energy changes with distance, rather than the energy itself, determines the variation of the Lorentz factor.
In its original form, the model's limit to the Lorentz factor growth is $\Gamma_\mathrm{sh} \propto R^{3/4}$. 
However, these works contain a tricky flaw. The authors take the change of the entire comoving volume of the shocked gas ($\propto R^3/\Gamma_\mathrm{sh}$), to measure the comoving volume change of an individual fluid element. But, along the radial flow lines, the comoving volume of a fluid element scales as $R^2 \Gamma_\mathrm{sh}$ (from the relativistic continuity equation). This  alters the anticipated blast wave dynamics. With the correct expression, this model leads to    $\Gamma_\mathrm{sh} \propto R$, which is  the common acceleration law of a hot fireball. 
Even this acceleration is too slow to explain the fast rise in GRB~221009A.

\begin{figure}
    \centering
    \includegraphics[width=0.8\linewidth]{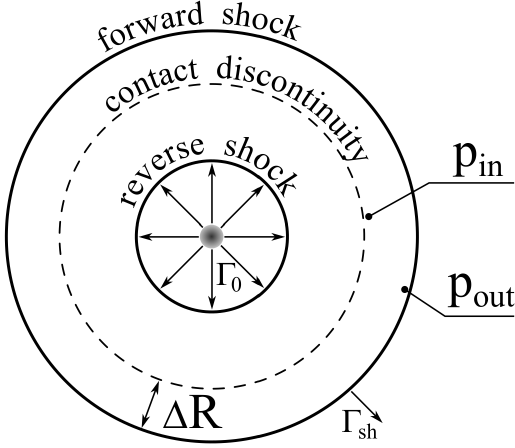}
    \caption{Schematic representation of the blast wave model. The shocked external material occupies the region between the contact discontinuity and the forward shock, whose width is $\Delta R \sim R/\Gamma_\mathrm{sh}^2 \ll R$. The region between the contact discontinuity and the reverse shock is filled with shocked ejected material. Note that the model treats the region between the forward shock and the reverse shock as infinitely thin, yet it takes into account the pressure difference across $\Delta R$.}
    \label{fig:ShockModelCartoon}
\end{figure}

We adopt a different approach. {Noting} that in GRB~221009A the afterglow was caught in its rising phase when there is an energy supply from the central engine and the blast wave still builds up,
we recall that at this stage, the system is {enclosed between two shocks} \citep{Sari1995} --- a forward shock that propagates into the external medium and a reverse shock that propagates into the ejecta (illustrated in Fig.~\ref{fig:ShockModelCartoon}). A contact discontinuity separates the shocked external matter and the shocked ejecta\footnote{The whole regions of shocked external matter and shocked ejecta are, at times, called the ``forward shock" and the ``reverse shock" colloquially.
We call them ``regions" to stress the fact that we deal with energy density, pressure, pressure gradient, etc., averaged over these regions rather than the quantities at the shocks themselves.}.
Energy is supplied to the blast wave from the jet {through} the reverse shock. As the reverse shock and the forward shock regions are connected via a contact 
discontinuity, the internal energy of the reverse shock region lost in its adiabatic expansion is transferred to the forward shock region.
The forward shock region guides the evolution of the blast wave's Lorentz factor. Unlike other models, we explicitly take into account the inertia of the shocked material so that the acceleration (or deceleration) rate is given by the average pressure gradient acting on the swept mass.

We provide here a simple analytic model for this system. While this model ignores the internal dynamics within the two shocked regions, such as {possible} appearance of additional shocks within them, the model fits nicely, as we show here, the observations of GRB 221009A. 
Our approach essentially averages the hydrodynamic equation of motion (in Lagrangian formulation) over the shocked {external} material. A fluid element in its locally comoving frame experiences acceleration $a = - c^2 \nabla p/w$, where $p$ is pressure and $w$ specific enthalpy ($w=4p$ for a relativistic equation of state). Then it's Lorentz factor, measured in the lab frame, changes as $\diff{\Gamma_\mathrm{el}}/\diff{R} = a/c^2$.

Averaging over the external shocked material, we obtain a differential equation that describes the evolution of the forward shock Lorentz factor (which is $\sqrt{2}$ times larger than the bulk Lorentz factor of the shocked gas):
\begin{equation}  \label{shock_acceleration_initial}
    \frac{\diff{\Gamma_\mathrm{sh}}}{\diff{R}} 
    = - \sqrt{2} \frac{\langle \nabla p \rangle}{\langle w \rangle}
    = \sqrt{2} \left( \frac{p_\mathrm{in}}{p_\mathrm{out}} - 1 \right) \frac{1}{\Delta R^{\prime}} \frac{p_\mathrm{out}}{\langle w \rangle} \, .
\end{equation}
Here, $\Delta R^{\prime}$ is the width of the forward shock zone in the comoving frame.

The average enthalpy $\langle w \rangle$ can be expressed in terms of the energy of shocked external material $E_\mathrm{fs}$
\begin{equation}  \label{avg_enthalpy}
    E_\mathrm{fs} = 4\pi R^2 \Delta R^{\prime} \langle w \rangle \frac{\Gamma_\mathrm{sh}}{\sqrt{2}} 
    \, .
\end{equation}
The pressure at the outer 
boundary is determined from the jump conditions at the shock,
\begin{equation}  \label{p_out}
    p_\mathrm{out} = \frac{2}{3} \Gamma_\mathrm{sh}^2 \rho c^2  \, .
\end{equation}

Substituting Eqs.~(\ref{avg_enthalpy}), (\ref{p_out}), and (\ref{rho_Msw_relation}) into Eq.~(\ref{shock_acceleration_initial}) we obtain
\begin{equation}  \label{shock_acceleration_interm}
    \frac{R}{\Gamma_\mathrm{sh}} \frac{\diff{\Gamma_\mathrm{sh}}}{\diff{R}} 
    = \left( \frac{p_\mathrm{in}}{p_\mathrm{out}} -1 \right) \frac{2 (3-k) \Gamma_\mathrm{sh}^2 M_\mathrm{fs} c^2}{3 E_\mathrm{fs}}    
    \, .
\end{equation}
When the energy of shocked external material is equal to the energy of the self-similar BM solution, i.e. $E_\mathrm{fs} = E_\mathrm{_{BM}} \equiv C_\mathrm{_E} \Gamma_\mathrm{sh}^2 M_\mathrm{fs} c^2$, then Eq.~(\ref{shock_acceleration_interm}) must reproduce the behaviour of this self-similar solution, i.e. 
$\left( R \middle/ \Gamma_\mathrm{sh} \right) \left( \diff{\Gamma_\mathrm{sh}} \middle/ \diff{R} \right) = (k-3)/2$. This constrain allows us to derive
\begin{equation}  \label{pressure_ratio_BM}
    \frac{p_\mathrm{in}}{p_\mathrm{out}}\Biggr|_\mathrm{_{BM}} = 1 - \frac{3}{4} C_\mathrm{_E}  \, .
\end{equation}

The pressure ratio $p_\mathrm{in}/p_\mathrm{out}$ for a general solution is different from Eq.~(\ref{pressure_ratio_BM}). To estimate it, we note that
$E_\mathrm{fs} \propto (p_\mathrm{in}+p_\mathrm{out})$ and $p_\mathrm{out}$ depends only on the local density and Lorentz factor of the forward shock. 
Then we calculate the ratio $p_\mathrm{in}/p_\mathrm{out}$ for any value of $E_\mathrm{fs}$ from the 
relation 
\begin{multline}  \label{pressure_ratio}
    E_\mathrm{fs} \left( \frac{p_\mathrm{in}}{p_\mathrm{out}} + 1 \right)^{-1}
    = E_\mathrm{_{BM}} \left( \frac{p_\mathrm{in}}{p_\mathrm{out}}\Biggr|_\mathrm{_{BM}} + 1 \right)^{-1} \\
    \quad \Rightarrow \quad
    \frac{p_\mathrm{in}}{p_\mathrm{out}} 
    = \frac{E_\mathrm{fs}}{E_\mathrm{_{BM}}} \left( 2 - \frac{3}{4} C_\mathrm{_E} \right) - 1 \, . \\
\end{multline}

We combine Eqs.~(\ref{shock_acceleration_interm}), (\ref{pressure_ratio}) to obtain the equation that describes evolution of the forward shock Lorentz factor in its final form:
\begin{equation}  \label{shock_acceleration_final}
    \frac{R}{\Gamma_\mathrm{sh}} \frac{\diff{\Gamma_\mathrm{sh}}}{\diff{R}} 
    =  C_\mathrm{_A} \left( 1 - \frac{E_\mathrm{_{BM}}}{E_\mathrm{fs}} \right) -  \frac{3-k}{2}  
    \, , \quad     
    C_\mathrm{_A} =  \frac{4 (3-k)}{3 \ C_\mathrm{_E}}     \, .
\end{equation}
In the case of ISM circumburst medium $C_\mathrm{_A}=34/3$ and in the case of wind-like circumburst medium $C_\mathrm{_A}=6$, so that Eq.~(\ref{shock_acceleration_final}) allows for a very rapid increase of the Lorentz factor, explaining the  observed $t_\mathrm{rise} \ll t_\mathrm{decay}$. .

The 
equality in Eq.~(\ref{shock_acceleration_final}) is approximate if the solution is not exactly self-similar, and hence, we consider $C_\mathrm{_A}$ as a semi-empirical adjustable parameter. The best-fit model may require a value of $C_\mathrm{_A}$ that is somewhat different from the value given in Eq.~(\ref{shock_acceleration_final}). 
When fitting GRB~221009A (see Sect.~\ref{sec:best_fit}), we  vary this coefficient and find that the calculated lightcurve is rather sensitive to the adopted value of $C_\mathrm{_A}$.

To estimate $E_\mathrm{fs}$ we recall that energy transferred from the free coasting jet is transferred first to the shocked ejecta matter. This region has a kinetic energy $\Gamma_\mathrm{sh} M_\mathrm{se} c^2$ and an internal energy $E_\mathrm{se}$. As the blast wave propagates, the expanding shocked ejecta  does work on the adjacent region of the shocked external gas. In this process, the internal energy of the shocked ejecta, $E_\mathrm{se}$, 
decreases, and the internal energy of the shocked swept material, $E_\mathrm{fs}$, increases by the same amount.
We calculate the internal energy change through the relativistic adiabatic equation, $p V^{4/3} = \text{const}$, together with expressions for internal energy (enthalpy), $E_\mathrm{int} = 4pV$, and for the comoving volume, $V \propto R^2 \Gamma_\mathrm{sh}$.
Consequently, the equations for shocked ejecta 's mass, $M_\mathrm{se}$, and internal energy are
\begin{equation}  \label{piston_mass_eq}
    \frac{\diff{M_\mathrm{se}}}{\diff{R}} =  \frac{L_\mathrm{kin}}{\Gamma_0 c^2} \frac{\diff{t_\mathrm{inj}}}{\diff{R}}
\end{equation}
and
\begin{equation}  \label{piston_energy_eq}
    \frac{\diff{E_\mathrm{se}}}{\diff{R}} =  
    \left( 1 - \frac{\Gamma_\mathrm{sh}}{\Gamma_0} \right) \left( 1 - \epsilon_\mathrm{r}^\mathrm{(rs)} \right)
    L_\mathrm{kin} \frac{\diff{t_\mathrm{inj}}}{\diff{R}}
    - \frac{1}{3} E_\mathrm{se} \left( \frac{2}{R} + \frac{1}{\Gamma_\mathrm{sh}} \frac{\diff{\Gamma_\mathrm{sh}}}{\diff{R}} \right)  .
\end{equation}
We take $\Gamma_\mathrm{sh}$ for the shocked ejecta Lorentz factor. In an exact hydrodynamic solution, it is approximately equal to the Lorentz factor of material at the contact discontinuity, which can be estimated through the Lorentz factor of shocked material immediately behind the forward shock, $\Gamma_\mathrm{sh}/\sqrt{2}$.
However, an attempt to account for the difference of the shocked ejecta Lorentz factor from $\Gamma_\mathrm{sh}$ would be fundamentally inconsistent with our treatment of the whole blast wave as a single object with zero width.

The first term in Eq.~(\ref{piston_energy_eq}) describes energy injection from the central engine (minus energy associated with the rest mass of the reverse shock region, and minus energy radiated at the reverse shock). The second term describes the adiabatic losses. In our model for GRB~221009A we neglect  radiative losses at the reverse shock, setting its radiative efficiency $\epsilon_\mathrm{r}^\mathrm{(rs)}$ to zero. This assumption is subsequently justified by the best-fit lightcurve model (see Sect.~\ref{sec:residuals}).

The energy lost in the adiabatic expansion of the shocked ejecta   
is transferred to the forward shock zone, and it is partially radiated. From the energy balance equation for the entire blast wave (see Eq.~\ref{energy_input}),

\begin{multline} 
\label{blast_wave_energy_balance}
    \Gamma_\mathrm{sh} M_\mathrm{se} c^2 + E_\mathrm{se} + E_\mathrm{fs} 
    = \int_0^{t_\mathrm{inj}} L_\mathrm{kin} \diff{t} \\
    - 2\pi \int_0^{R} \epsilon_\mathrm{r} \Gamma_\mathrm{sh}^2 \rho c^2 \left( R^{\prime} \right)^2 \diff{R^{\prime}}  
    - \int_0^{t_\mathrm{inj}}  \epsilon_\mathrm{r}^\mathrm{(rs)} \left( 1 - \frac{\Gamma_\mathrm{sh}}{\Gamma_0} \right)  
    L_\mathrm{kin} \diff{t}     \, .
\end{multline}
Here the last term is the energy radiated from the reverse shock.
Substituting into it Eqs.~(\ref{piston_mass_eq}), we calculate the energy of the forward shock zone
\begin{multline} \label{fs_energy_eq}
    E_\mathrm{fs} 
    = \left( 1 - \frac{\Gamma_\mathrm{sh}}{\Gamma_0} \right)  \int_0^{t_\mathrm{inj}} L_\mathrm{kin} \diff{t}
    - \ E_\mathrm{se}   \\ 
    - 2\pi \int_0^{R} \epsilon_\mathrm{r} \Gamma_\mathrm{sh}^2 \rho c^2 \left( R^{\prime} \right)^2 \diff{R^{\prime}}  
    - \int_0^{t_\mathrm{inj}}  \epsilon_\mathrm{r}^\mathrm{(rs)} \left( 1 - \frac{\Gamma_\mathrm{sh}}{\Gamma_0} \right)  
    L_\mathrm{kin} \diff{t}     \, .
\end{multline}
Note an important difference between the integral equation for $E_\mathrm{fs}$ and the differential equation for $E_\mathrm{se}$: 
the first term in Eq.~(\ref{fs_energy_eq}) refers to local value of $\Gamma_\mathrm{sh}$ at a given distance, whereas 
the first term in Eq.~(\ref{piston_energy_eq}) and the last term in Eq.~(\ref{fs_energy_eq}) imply integrating $\Gamma_\mathrm{sh}$ over distance.

If $\Gamma_0 = \text{const}$, then the derivative of the injection time over the blast wave radius is simply
\begin{equation} \label{injection_time_derivative}
    \frac{\diff{t_\mathrm{inj}}}{\diff{R}} 
    = \frac{1}{2 c} \left( \frac{1}{\Gamma_\mathrm{sh}^2}  - \frac{1}{\Gamma_0^2} \right)  \, .
\end{equation}
If $\Gamma_0$ varies in time, then one has to solve Eq.~(\ref{injection_time}) first and then differentiate.
This expression completes the model's set of equations.

\subsection{Dimensionless equations of the model}
\label{sec:model_dimensionless}

For convenience, we summarize all the model's equations, 
rewriting them in a dimensionless form to reveal the model's independent parameters.
An afterglow is characterized by  two  dimensional and two dimensionless parameters.   The first two are the isotropic equivalent kinetic energy, $E^\mathrm{(iso)}_\mathrm{kin}$ and the 
dimensional deceleration time 
\begin{equation} \label{t_d}
    t_\mathrm{d} = R_\mathrm{d} / \left( 2 \Gamma_0^2 c \right) \ .
\end{equation}
The latter is related to deceleration distance $R_\mathrm{d}$ defined from
\begin{equation} \label{R_d}
    M_\mathrm{d} \equiv M_\mathrm{fs} \left(R_\mathrm{d}\right) 
    = \frac{4\pi R_\mathrm{d}^3}{3-k} \ \rho \left( R_\mathrm{d} \right) \ ,
\end{equation}
where
\begin{equation} \label{M_d}
    M_\mathrm{d} = \frac{E^\mathrm{(iso)}_\mathrm{kin}}{C_\mathrm{_E} \Gamma_0^2 c^2}
\end{equation}
is the deceleration mass.
{One may equivalently choose density as the second dimensional parameter, but this is less convenient when the density varies with distance. }

The two dimensionless parameters are the radiative efficiency
of the blast wave $\epsilon_\mathrm{r}$ (plus radiative efficiency of the reverse shock, $\epsilon_\mathrm{r}^\mathrm{(rs)}$, if needed)\footnote{
In our {simulations} we assume a constant radiative efficiency. This is consistent with GRB~221009A observations. At the same time, the equations that we derive are valid for any $\epsilon_\mathrm{r} (R)$ dependence.
Judging from our model for GRB~221009A, the radiative efficiency of the reverse shock can be neglected in models of blast wave dynamics.
}
and the reduced opening angle of the jet $\Gamma_0 \theta_j$, where $\theta_j$ is the effective opening angle of the jet. {We use a Gaussian jet profile, but other choices also work with the model.}
Neither $\theta_j$ nor $\Gamma_0$ is an independent parameter alone. The former only enters the lightcurve model through the reduced opening angle in combination with $\Gamma_0$, whereas the latter also enters the model's equations through $t_\mathrm{d}$, $R_\mathrm{d}$, and $M_\mathrm{d}$, but never explicitly. 
For simplicity, we consider the Lorentz factor of the ejecta $\Gamma_0$ as constant. Again, this   is consistent with  GRB~221009A observations.
In addition to the source's parameters, 
the parameter $k$ defines the density profile of the circumburst medium, $\rho \propto r^{-k}$, which can be either wind-like ($k=2$) or a constant density ($k=0$). 
Finally, $C_\mathrm{_A}$ is  a model's adjustable factor, for which   we test in our fitting procedure  values around the estimate given in Eq.~(\ref{shock_acceleration_final}).

We introduce the following dimensionless variables: \\
The dimensionless energies 
\begin{equation} \label{dimensionless_energy}
    {\cal E}_\mathrm{p,fs} = E_\mathrm{p,fs}/E^\mathrm{(iso)}_\mathrm{kin} \ .
\end{equation}
A dimensionless time
\begin{equation} \label{dimensionless_time}
    \tau = t_\mathrm{inj}/t_\mathrm{d} \ .
\end{equation}
A normalized  source's kinetic power 
\begin{equation}  \label{dimensionless_power}
    {\cal L} =  \frac{L_\mathrm{kin} t_\mathrm{d}}{E^\mathrm{(iso)}_\mathrm{kin}}    \, .
\end{equation}
A dimensionless blast wave radius
\begin{equation} \label{dimensionless_diatance}
    r = R/R_\mathrm{d} \ .
\end{equation}
A dimensionless swept mass
\begin{equation} \label{dimensionless_mass}
    m = M_\mathrm{fs}/M_\mathrm{d} \ .
\end{equation}
We will also use a normalized Lorentz factor
\begin{equation} \label{reduced_LF}
    \gamma = \Gamma_\mathrm{sh}/\Gamma_0 \ .
\end{equation}

The dimensionless equations of  our blast wave model are:  \\
The injection time equation (from Eq.~\ref{injection_time_derivative})
\begin{equation} \label{dimensionless_injection_time_eq}
    \frac{\diff{\tau}}{\diff{r}}  =   \frac{1}{\gamma^2}  - 1  \, .
\end{equation}
The shocked ejecta energy equation (from Eq.~\ref{piston_energy_eq})
\begin{equation}  \label{dimensionless_piston_energy_eq}
    \frac{\diff{{\cal E}_\mathrm{se}}}{\diff{r}} 
    =  \left( 1 - \gamma \right)  \left( 1 - \epsilon_\mathrm{r}^\mathrm{(rs)} \right) {\cal L}  \frac{\diff{\tau}}{\diff{r}}  
    - \frac{1}{3} {\cal E}_\mathrm{se}  \left( \frac{2}{r} + \frac{1}{\gamma} \frac{\diff{\gamma}}{\diff{r}} \right)  
    \ .
\end{equation}
The forward shock zone energy equation (from Eq.~\ref{fs_energy_eq})
\begin{multline} \label{dimensionless_fs_energy_eq}    
    {\cal E}_\mathrm{fs} 
    = \left( 1 - \gamma \right) \int_0^{\tau} \left[ 1 - \epsilon_\mathrm{r}^\mathrm{(rs)} \left( 1 - \gamma \right)  \right]
    {\cal L} \diff{\tau^{\prime}}  \\
    -  \int_0^{r} \bar{\epsilon}_\mathrm{r} \left( r^{\prime} \right)^{2-k} \gamma^2 \diff{r^{\prime}}
    - {\cal E}_\mathrm{se}   
    \, , \quad
    \bar{\epsilon}_\mathrm{r} = \frac{3-k}{2 \ C_\mathrm{_E}} \ \epsilon_\mathrm{r}     \, .
\end{multline}
The blast wave Lorentz factor equation (from Eq.~\ref{shock_acceleration_final})
\begin{equation}  \label{dimensionless_LF_evolution_eq}
    \frac{r}{\gamma} \frac{\diff{\gamma}}{\diff{r}} 
    =  C_\mathrm{_A} \left( 1 - \frac{\gamma^2  m}{{\cal E}_\mathrm{fs}} \right) -  \frac{3-k}{2}  
    \, , \quad     
    C_\mathrm{_A} \approx  \frac{4 (3-k)}{3 \ C_\mathrm{_E}}     \, .
\end{equation}
A dimensionless swept mass - radius relation  
\begin{equation}  \label{dimensionless_mass_radius_ralation}
    m = r^{3-k}  \, .
\end{equation}
Equation~(\ref{piston_mass_eq}) for the reverse shock region mass is auxiliary. 

The jet's kinetic power $L_\mathrm{kin}(t)$ is not directly observed. Our assumption throughout this paper is that the released kinetic energy is  proportional to the GRB's prompt luminosity.

\subsection{Application to narrow jets}
\label{sec:model_jet}

The model that we construct here is one-dimensional. Although the immediate application is for spherically symmetric explosions, it also works without any change in a situation when the flow lines are exactly radial. This is a good approximation if the blast wave's angular scale (set by the jet's opening angle $\theta_j$) is much larger than $1/\Gamma_\mathrm{sh}$, because the flow lines separated by an angle larger than $1/\Gamma$ are causally disconnected. All one needs in this case is to calculate the model's dimensional parameters ($E^\mathrm{(iso)}_\mathrm{kin}$, $t_\mathrm{d}$, $R_\mathrm{d}$, and $M_\mathrm{d}$) as functions of angle in accordance to the jet's angular profile, and then solve the equations independently for each direction.

A blast wave that has decelerated to $\Gamma_\mathrm{sh} < 1/\theta_j$ has to be treated with caution. It has been suggested that in this case, the shocked material may expand sideways with near-sonic velocity, so that at later times the blast wave keeps its opening angle approximately equal to $1/\Gamma_\mathrm{sh}$ and decelerates exponentially with distance \citep[e.g., ][]{Rhoads1999,Sari1999}. 
This theoretical conclusion was employed to explain some observations where afterglow lightcurves become steeper after some time. Such lightcurve breaks are now commonly considered as a manifestation of change in the blast wave's deceleration law at the moment when $\Gamma_\mathrm{sh} \sim 1/\theta_j$, hence the name --- jet breaks.
Subsequent numerical simulations, however, showed a much slower than anticipated lateral evolution \citep[e.g., ][]{KumarGranot2003}. This is supported by several analytic works \citep[see e.g.][]{GranotPiran2012,GS-Nakar2023}.

No matter when exactly the lateral expansion starts to dominate the blast wave's deceleration law, the jet break in afterglow lightcurves always occurs when $\Gamma_\mathrm{sh} \sim 1/\theta_j$. Even in the absence of sideways expansion, starting from this moment, an observer sees the blast wave filling only a part of the visible zone, which is restricted to the cone with opening angle $\sim 1/\Gamma_\mathrm{sh}$ around the line of sight (the outer parts appear negligibly faint because of rapidly declining Doppler factor).
So, the apparent luminosity (and the equivalent isotropic energy) is smaller compared to a spherical blast wave. 

We find (see Sect.~\ref{sec:best_fit}) that the observed GRB~221009A lightcurve is best-fitted with a narrow jet whose opening angle is $\theta_j \sim 1/\Gamma_0$. Moreover, our model, which assumes radial flow lines and thus ignores lateral spreading of the blast wave, demonstrates very good agreement with observations. This suggests that relativistic blast waves are indeed not so prone to lateral expansion and may preserve their original angular structure even after they decelerate to Lorentz factors less than $1/\theta_j$.

\subsection{The different types of solutions}
\label{sec:model_examples}

To explore the possible lightcurves expected with our model, 
we consider a GRB whose prompt emission has a single pulse of triangular shape and a total duration of $t_\mathrm{_{GRB}}$. Again, we assume that the jet's kinetic power is proportional to the prompt luminosity, and we calculate the afterglow lightcurves for six qualitatively different scenarios. These include: 
very slowly decelerating blast wave with $t_\mathrm{d} = 100 t_\mathrm{_{GRB}}$ (we will call this situation a lagging afterglow), 
a rapidly  decelerating blast wave with $t_\mathrm{d} = 0.01 t_\mathrm{_{GRB}}$ (we will call this situation fast afterglow), 
and an intermediate case with $t_\mathrm{_{GRB}} = t_\mathrm{d}$.
We repeat this for wind-like circumburst density profiles (we will call these lightcurves wind-type) and for  constant-density circumburst media (ISM-type lightcurves). 
In all cases, we compare lightcurves from narrow on-axis jets with $\Gamma_0 \theta_j = 3$ 
to lightcurves from spherically symmetric outflows (or very wide jets).

Our model lightcurves are bolometric, calculated with the theoretical value for the coefficient $C_\mathrm{_A}$ (see Eq.~\ref{shock_acceleration_final}). In this section, we don't take into account photon-photon annihilation, which is included in the models fitted later to GRB 221009A. 

\begin{figure}
    \centering
    \includegraphics[width=1.0\linewidth]{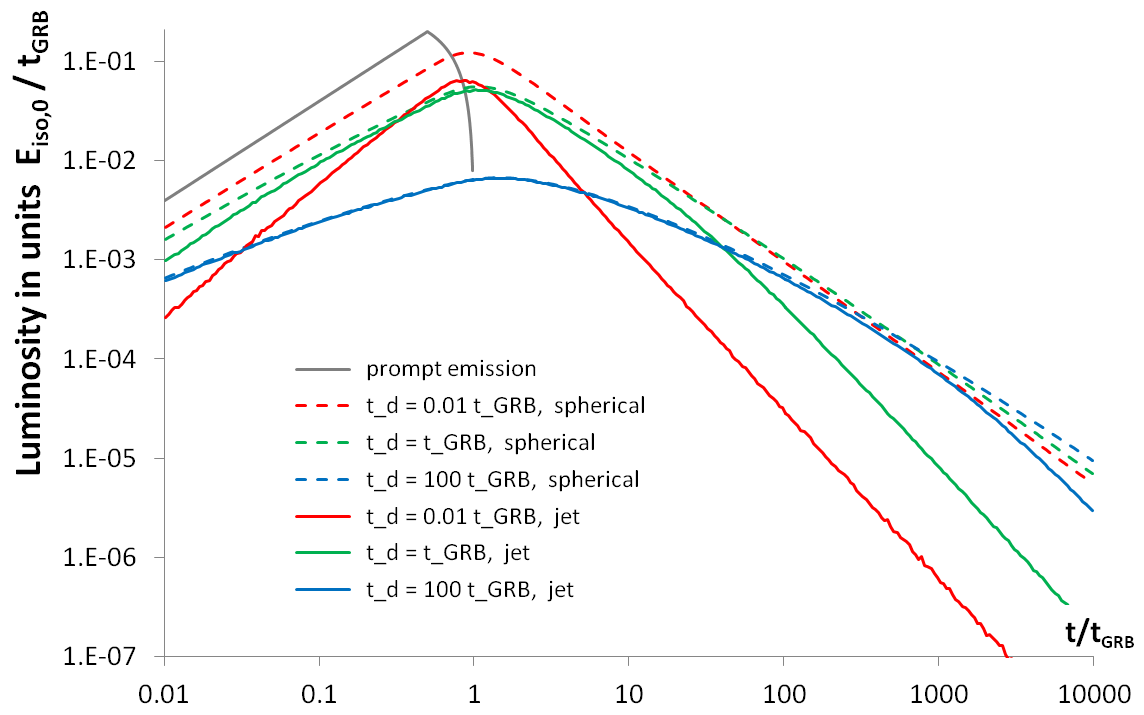}
    \caption{ 
    Wind-type lightcurves for afterglows with different $t_\mathrm{d}/t_\mathrm{_{GRB}}$ ratios.
    The prompt luminosity is scaled down by factor $0.1$. 
    In all models, the radiative efficiency is $\epsilon_\mathrm{r}=0.2$.
    Models with jetted outflows are for the Gaussian jet profile with a reduced width $\Gamma_0 \theta_j = 3$ and an on-axis line of sight.
    $E_\mathrm{iso,0}$ is the isotropic equivalent energy on the jet's axis.
    }
    \label{fig:wind-type_lightcurves}
\end{figure}

Figure~(\ref{fig:wind-type_lightcurves}) shows wind-type lightcurves. 
The most remarkable feature is that 
all these lightcurves peak at $t \approx t_\mathrm{_{GRB}}$, and the tendency to increase the peak time with increasing $t_\mathrm{d}$ is almost unnoticeable. However, the ratio $t_\mathrm{d}/t_\mathrm{_{GRB}}$ significantly influences the shape of the peak:
lagging
afterglows have very broad peaks, whereas fast afterglows have relatively sharp peaks.
In all cases, the shape of the prompt pulse has little influence on the afterglow's lightcurves.
Note that 
{in jet models, the lightcurves not only may  decay faster compared to lightcurves for spherically symmetric models, but} 
may also have a faster rise. This is a manifestation of the same effect as in jet breaks --- when the shock's Lorentz factor is smaller than $1/\theta_j$, then the blast wave does not fill the entire cone potentially visible to the observer.
This reduces the averaged over visible region shock power and decreases the apparent brightness. Dimming due to 
this effect becomes unimportant when the {\it accelerating} blast wave starts to fulfill the condition $\Gamma_\mathrm{sh} \gtrsim 1/\theta_j$, after this moment, the lightcurve's behaviour is similar to the isotropic case. 
We will call this phenomenon the inverse jet break.

\begin{figure}
    \centering
    \includegraphics[width=1.0\linewidth]{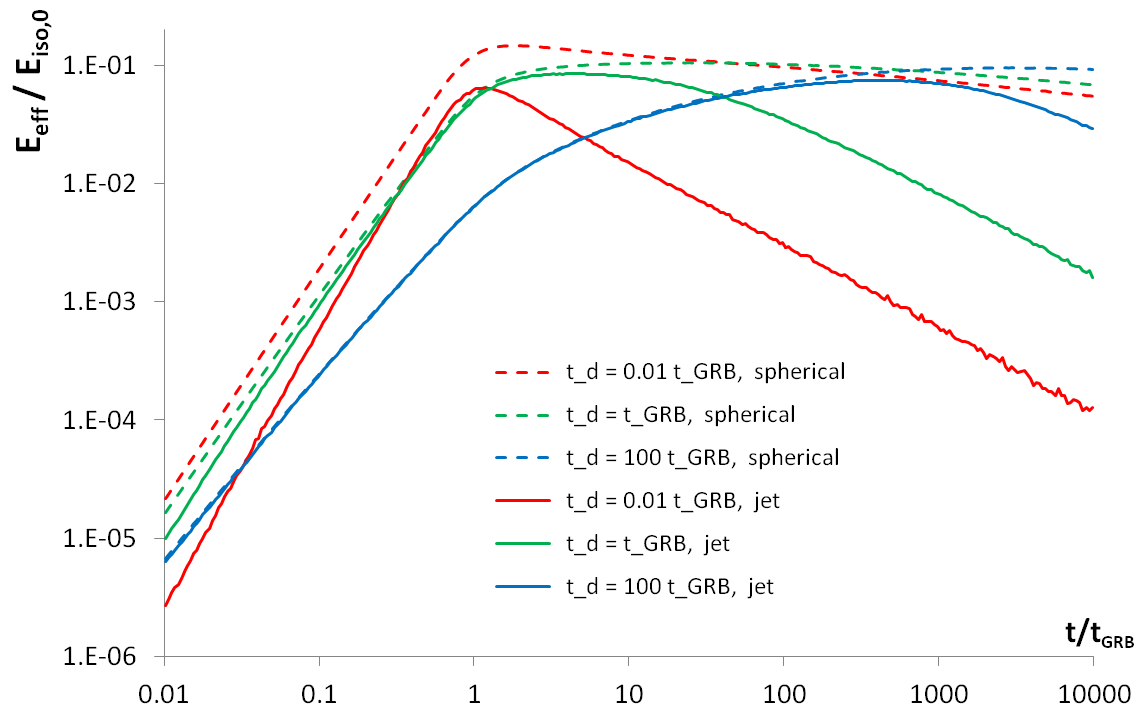}
    \caption{ 
    Same models as in Fig.~(\ref{fig:wind-type_lightcurves}) presented as plots of the efficient energy $E_\mathrm{eff} = L t$.
    }
    \label{fig:wind-type_Eeff}
\end{figure}

The effective energy plots for the same models (see Fig.~\ref{fig:wind-type_Eeff}) show a qualitative difference between them. In particular, intermediate and lagging
afterglows have a very broad maximum/plateau phase, which is shifted to a later time compared to the lightcurve's peak.


\begin{figure}
    \centering
    \includegraphics[width=1.0\linewidth]{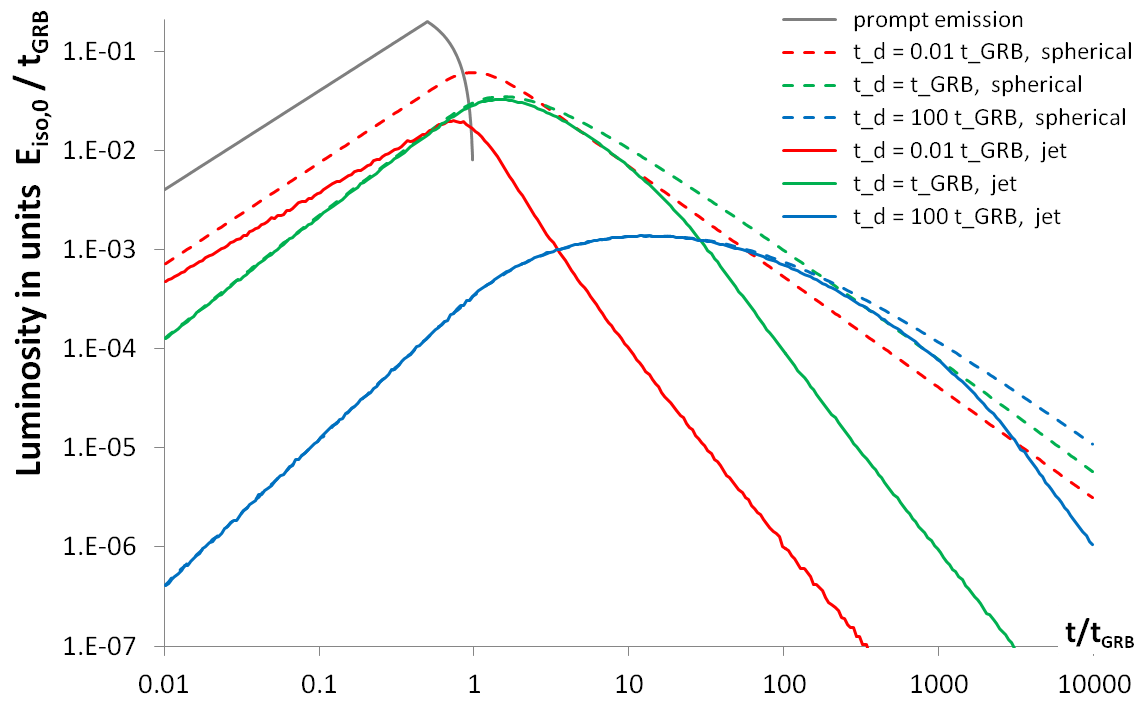}
    \caption{ 
    ISM-type lightcurves for afterglows with different $t_\mathrm{d}/t_\mathrm{_{GRB}}$ ratios.
    The prompt luminosity is scaled down by a factor $0.1$.
    In all models the radiative efficiency is $\epsilon_\mathrm{r}=0.2$.
    Models with jetted outflows are for  Gaussian jet profiles with  a reduced width $\Gamma_0 \theta_j = 3$ and an on-axis line of sight.
    $E_\mathrm{iso,0}$ is the isotropic equivalent energy on the jet's axis.
    }
    \label{fig:ISM-type_lightcurves}
\end{figure}

\begin{figure}
    \centering
    \includegraphics[width=1.0\linewidth]{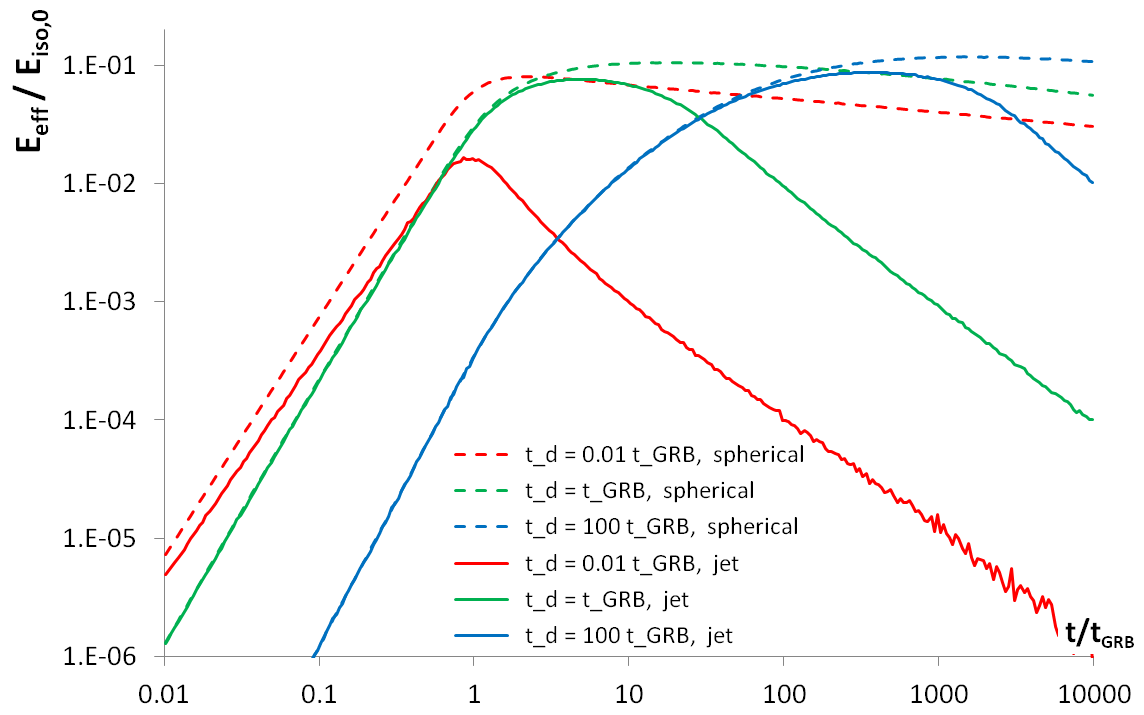}
    \caption{ 
    Same models as in Fig.~(\ref{fig:ISM-type_lightcurves}) presented as plots of the efficient energy $E_\mathrm{eff} = L t$.
    }
    \label{fig:ISM-type_Eeff}
\end{figure}

Figure~(\ref{fig:ISM-type_lightcurves}) shows ISM-type lightcurves. Unlike the wind-type lightcurves, the peak location depends on the blast wave deceleration time and can be estimated as $\approx \min(t_\mathrm{_{GRB}}, 0.1 t_\mathrm{d})$. Overall, the shape of  the lightcurves' peaks follows the same pattern as for wind-like lightcurves --- fast afterglows have sharp peaks and lagging
afterglows have broad ones, though they never become as broad as in wind-type lightcurves.
In the effective energy plots (see Fig.~\ref{fig:ISM-type_Eeff}), we again see the formation of a very broad maximum/plateau phase, for lagging afterglows. However, in the ISM-type solutions,  and this is an important difference from the wind-type solutions,   the plateau phase shifts to a later time together with the lightcurve's peak. In other words, if the plateau phase is present, then both its beginning and its end are delayed with respect to the peak of the prompt emission.


\begin{figure}
    \centering
    \includegraphics[width=1.0\linewidth]{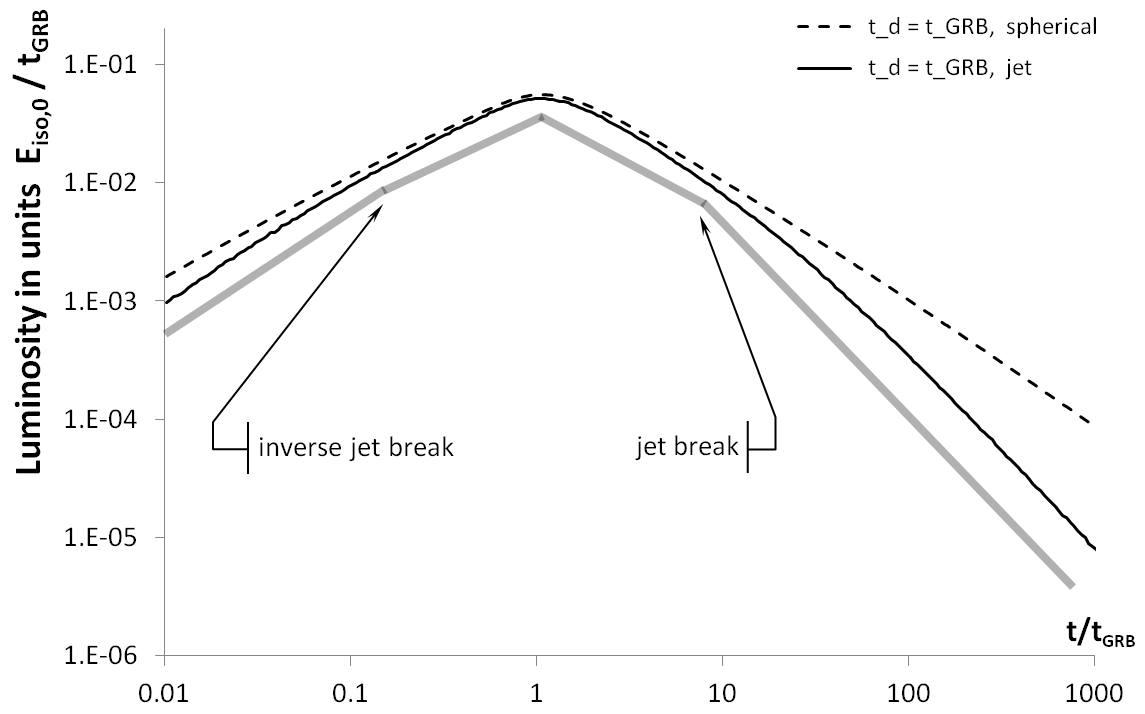}
    \caption{ 
    Comparison of simulated wind-type bolometric lightcurves for spherical blast wave and for a jet with reduced opening angle $\Gamma_0 \theta_j = 3$.
    The thick gray polygonal chain follows the lightcurve for the jet solution and highlights its main features.
    }
    \label{fig:AnnotatedLightcurve}
\end{figure}

Figure~(\ref{fig:AnnotatedLightcurve}) presents separately a wind-type lightcurve that illustrates all features that we expect: the inverse jet break at an early time, then the peak, then the (regular) jet break.

As expected, the jet breaks are more pronounced in the ISM-type lightcurves.
The inverse jet breaks are more pronounced in the wind-type solutions. 
Indeed, for an inverse jet break to appear there must be a phase of blast wave acceleration, and it readily occurs in wind-like density profiles.
The existence of such a phase in a constant-density surrounding is possible only if the central engine's power as a function of time satisfies some conditions.
Note that very narrow jets may have their inverse jet break coincident with the lightcurve's peak.

\section{Theoretical fit to GRB~221009A lightcurve} 
\label{sec:lightcurve_fit}

\subsection{Lightcurve computation } 
\label{sec:method}

We calculate jet's kinetic power under the assumption that it is proportional to the prompt luminosity, i.e. $L_\mathrm{kin} = \left(1/\eta_\mathrm{rad}- 1 \right) L_\mathrm{pr}$, where $\eta_\mathrm{rad}$ is the radiative efficiency at the prompt phase. In practice, we use the count rate shown in Fig.~\ref{fig:Xray_lightcurve} to measure $L_\mathrm{pr}$ and hence $L_\mathrm{kin}$.

The lightcurves that we calculate are bolometric, and we take into account photon-photon annihilation that arises from interaction of afterglow photons with prompt photons. The jet is structured with a constant Lorentz factor but variable  kinetic power as a function of $\theta$. The blast wave follows the angular profile of the jet, that  we assume  to be Gaussian.

To compute a model lightcurve we split the blast wave into many small segments, each of them subtending an angle $\theta_\mathrm{seg} \ll \theta_j$. Then we sum signals from all segments to obtain lightcurve for the whole blast wave. 

For every segment, we calculate $\Gamma_\mathrm{sh} (R,\theta)$ dependence along the flow line that goes through the center of this segment at angle $\theta$ to the jet's axis. To do this we solve dimensionless equations (\ref{dimensionless_injection_time_eq}, \ref{dimensionless_piston_energy_eq}, \ref{dimensionless_fs_energy_eq}, \ref{dimensionless_LF_evolution_eq}, and \ref{dimensionless_mass_radius_ralation}), thus we in effect ignore sideways expansion of the blast wave. 
Dimensional quantities are recovered by applying angle-dependent scale factors according to 
\begin{equation}
\begin{array}{rl}
    E^\mathrm{(iso)}_\mathrm{kin}\!\left( \theta \right) =& J\!\left( \theta \right) E^\mathrm{(iso)}_\mathrm{kin}\!\left( 0 \right) \ , \\
    \Gamma_0\!\left( \theta \right) =& \Gamma_0\!\left( 0 \right) \ , \\
    M_\mathrm{d}\!\left( \theta \right) =& J\!\left( \theta \right) M_\mathrm{d}\!\left( 0 \right) \ , \\
    R_\mathrm{d}\!\left( \theta \right) =& \left[ J\!\left( \theta \right) \right]^{\frac{1}{3-k}} R_\mathrm{d}\!\left( 0 \right) \ , \\
    t_\mathrm{d}\!\left( \theta \right) =& \left[ J\!\left( \theta \right) \right]^{\frac{1}{3-k}} t_\mathrm{d}\!\left( 0 \right) \ . \\
\end{array}
\end{equation}
We consider a Gaussian jet profile 
\begin{equation} \label{jet_profile}
    J\!\left( \theta \right) = \exp \left( -\theta^2/ \theta_j^2 \right) 
\end{equation}
and a jet's axis aligned with the line of sight. An off-axis geometry does not result in a measurable improvement of the lightcurve fit, but it increases the already large central engine's power estimate. 
We assume a constant and independent on the angle jet's Lorentz factor,  and the same temporal profile for $L_\mathrm{kin}\!\left( t \right)$ in all directions, which is to be rescaled into dimensionless time to obtain ${\cal L}\!\left( \tau \right)$ that enters the equations.

\begin{figure}
    \centering
    \includegraphics[width=1.0\linewidth]{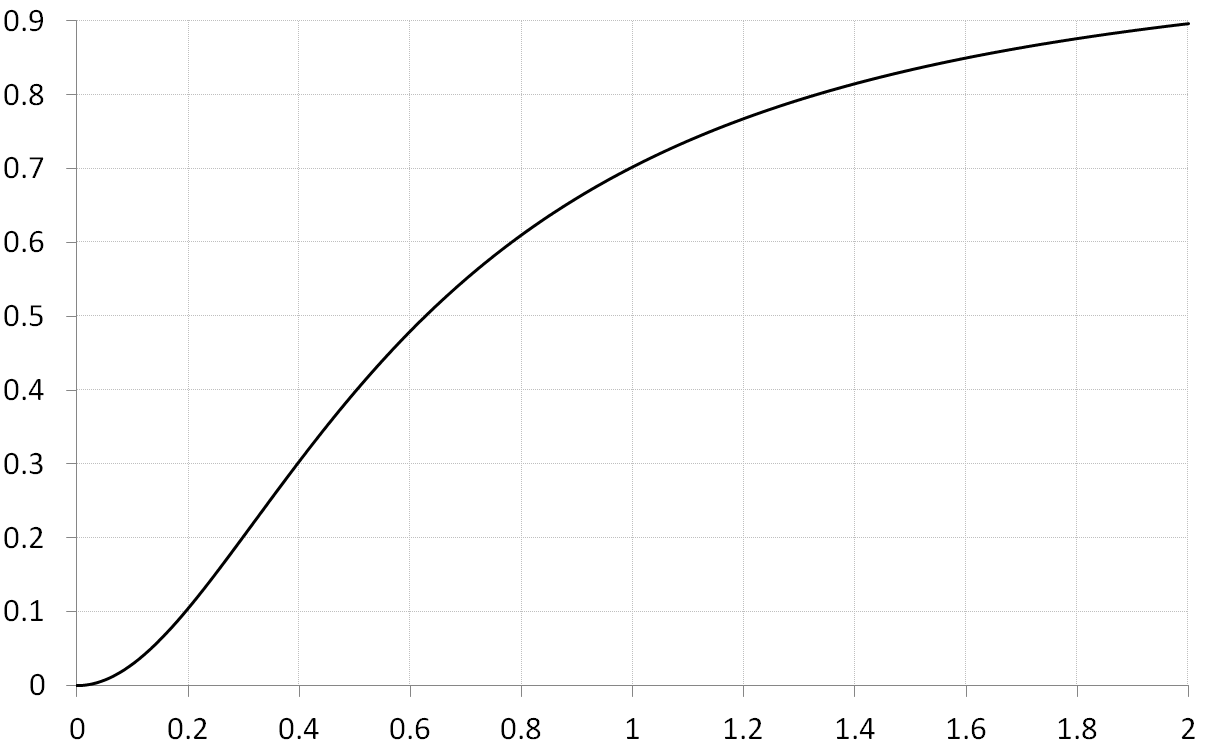}
    \caption{
    The jet width correction factor $f_\mathrm{w}$,  defined as the ratio of the apparent luminosity of a jet observed on-axis to the isotropic equivalent luminosity calculated on the jet's axis (Eq.~\ref{correction_factor}), for a Gaussian jet profile (Eq.~\ref{jet_profile}) and in the ultrarelativistic limit.
    The difference of the function $f_\mathrm{w} (\Gamma_0,\theta_j)$ from its ultrarelativistic limit $f_\mathrm{w} (\Gamma_0 \theta_j)$ is $O\left( \Gamma_0^{-2} \right)$.
    }
    \label{fig:LuminosityCorrectionFactor}
\end{figure}

Since we do not limit our study to the case $\Gamma_0 \theta_j \gg 1$, we apply jet width correction factor to relate the apparent isotropic equivalent luminosity (and hence energy) to actual on-axis luminosity (energy):
\begin{equation}
    L^\mathrm{(iso)} = L^\mathrm{(iso)}_\mathrm{app} / f_\mathrm{w}  \ , \quad
    E^\mathrm{(iso)} = E^\mathrm{(iso)}_\mathrm{app} / f_\mathrm{w}  \ ,
\end{equation}
where
\begin{equation} \label{correction_factor}
    f_\mathrm{w} = 
    \left. 
    \int_0^{\pi} \frac{J\!\left( \theta \right) \sin \theta \, \diff{\theta}}{\Gamma_0^4 \left( 1- \beta_0 \cos \theta \right)^4}
    \middle/ 
    \int_0^{\pi} \frac{\sin \theta \, \diff{\theta}}{\Gamma_0^4 \left( 1- \beta_0 \cos \theta \right)^4}
    \right. \ .
\end{equation}
The function $f_\mathrm{w} (\Gamma_0,\theta_j)$ in the ultrarelativistic limit becomes the function of reduced jet width, $f_\mathrm{w} (\Gamma_0 \theta_j)$. It is plotted in Fig.~(\ref{fig:LuminosityCorrectionFactor}).

Once the hydrodynamic evolution of the blast wave is calculated along all necessary directions, we proceed with calculating the bolometric lightcurve. This is done by Monte Carlo method. Namely, at every timestep we generate a photon at a random, isotropically distributed direction in the shocked matter comoving frame and assign a weight that equals the comoving-frame energy of the shocked mass element times the radiative efficiency, i.e.,
\begin{equation}
    w = \epsilon_\mathrm{r} \frac{\Gamma_\mathrm{sh}}{\sqrt{2}}  \rho c^2 R^2 \diff{\Omega} \diff{R} 
      =  \frac{\epsilon_\mathrm{r} E^\mathrm{(iso)}_\mathrm{kin}}{\sqrt{2} C_\mathrm{_E} \Gamma_0} \gamma \diff{m} \ ,
\end{equation}
where $\diff{\Omega}$ is the solid angle subtended by the blast wave segment. 
In the next step, we apply a Lorentz boost into the lab frame (with Lorentz factor $\Gamma_\mathrm{sh}/\sqrt{2}$) 
and check if the photon moves at a small enough angle with respect to the line of sight (other photons are rejected). 
If the photon passes the direction check, its weight is multiplied by the Doppler factor, which derives from  Lorentz boost from the comoving frame into the lab frame. Finally, we propagate the photon from its place of  origin to the observer, calculating the probability of two-photon annihilation with another photon along its path. The main source of opacity is the prompt emission photons, for which we take the actual lightcurve (see Fig.~\ref{fig:Xray_lightcurve}) and use a spectral fit with a Band function suggested for the largest  prompt emission pulse ($\alpha = -0.76$, $\beta = -2.13$, $E_\mathrm{peak} = 3.038$~MeV, \citep[see][]{KonusWind}).
Unless stated otherwise, we propagate 1~TeV photons. Those of them that escape without interaction contribute to the lightcurve.

We fit the simulated lightcurves to the set of 9 reference points shown in the upper panel of Fig.~(\ref{fig:TeV_lightcurve}). 
The best fit is obtained by varying two parameters, $C_\mathrm{_A}$ and $t_\mathrm{d}$, for several values of the reduced jet width, $\Gamma_0 \theta_j$. In all simulations, we keep the same  radiative efficiency $\epsilon_\mathrm{r} = 0.2$.  We adjust the TeV radiative efficiency, $\epsilon_\mathrm{r}^\mathrm{_{(TeV)}}$, to match the observed TeV fluence, varying it between different simulations but  keeping it constant  throughout each  individual {simulation.} 
This assumption of a constant $\epsilon_\mathrm{r}^\mathrm{_{(TeV)}}$ is motivated by the lack of statistically significant spectral evolution in the TeV part of GRB~221009A spectrum \citep{LHAASO}. It is confirmed by the fact that this assumption leads  to a good lightcurve fit.

Calculating the energy injection rate we use the entire lightcurve from Fig.~(\ref{fig:Xray_lightcurve}), and all its parts are important. Even the very weak initial pulse at the trigger time (at $T_* - 226$~s) plays a role. By the beginning of the main phase of the central engine's activity, the blast wave started by the initial pulse already has a measurable mass. This slows blast wave acceleration and delays the afterglow's rise.

\subsection{The Best-fit Lightcurve} 
\label{sec:best_fit}

\subsubsection{Wind models}
\label{sec:winds}

Wind-type models produce very good fits to the data.
Table~\ref{tab:fit_parameters} summarizes the parameters of the best-fit lightcurve models  for several values of a reduced jet width. 
Good solutions not only exist but also show a degeneracy with respect to the reduced jet width. 
{All narrow jets produce similarly-looking solutions, where the main difference is at early time --- the smaller is jet width, the faster is the rise in the model lightcurve.
Narrower jets require a larger normalized efficiency of TeV emission and are closer to expectations in this respect.}
Wider jets with $\Gamma_0 \theta_j > 1$ result in progressively worse fits as the reduced jet width increases. 
The inverse jet break, discussed in Sect.~(\ref{sec:model_examples}) and seen better in the wind-type model solutions, contributes to the (a priori unexpected) fast rise of the lightcurve. 

\begin{table} 
\caption{
Parameters of the best-fit lightcurve models in a wind-like density profile for several values of the reduced jet width.
The asterisk denotes our choice for the reference model.
The value $\Gamma_\mathrm{sh}^\mathrm{(max)}$ is the maximal value of the blast wave's Lorentz factor during its evolution and
$\displaystyle X_\mathrm{_{TeV}} = \left( \frac{1}{\eta_\mathrm{rad}} - 1 \right) \frac{\epsilon_\mathrm{r}^\mathrm{_{(TeV)}}}{\epsilon_\mathrm{r}}$
is the normalized efficiency of TeV emission required to fit the observed TeV fluence.
}
\label{tab:fit_parameters}
\begin{tabular}{ccccccc}
\hline
    & $\Gamma_0 \theta_j$ & $C_\mathrm{_A}$ & $t_\mathrm{d}$, s 
    & $\Gamma_\mathrm{sh}^\mathrm{(max)} / \Gamma_0$  &  $\Gamma_\mathrm{sh}^\mathrm{(max)} \theta_j$ 
    &  $X_\mathrm{_{TeV}}$ \\
    \hline
    & 0.4 & 6.95 & 160~s & 0.71 & 0.28 &  0.091 \\
    \hline  
    & 0.5 & 7.0 & 145~s & 0.71 & 0.35 &  0.080 \\
    \hline  
    * & 0.6 & 7.0 & 130~s & 0.70 & 0.42 &  0.071 \\
    \hline  
    & 0.7 & 7.05 & 115~s & 0.69 & 0.49 &  0.064 \\
    \hline  
    & 0.8 & 7.0 & 100~s & 0.69 & 0.55 &  0.057 \\
    \hline  
    & 1.0 & 7.0 & 80~s & 0.68 & 0.68 &  0.048 \\
    \hline  
    & 1.2 & 6.95 & 60~s & 0.66 & 0.79 &  0.040 \\
    \hline  
    & 1.5 & 6.85 & 45~s & 0.65 & 0.97 &  0.034 \\
    \hline  
    & 2.0 & 6.7 & 30~s & 0.63 & 1.25 &  0.029 \\
    \hline  
\end{tabular}
\end{table}

\begin{figure}
    \centering
    \includegraphics[width=1.0\linewidth]{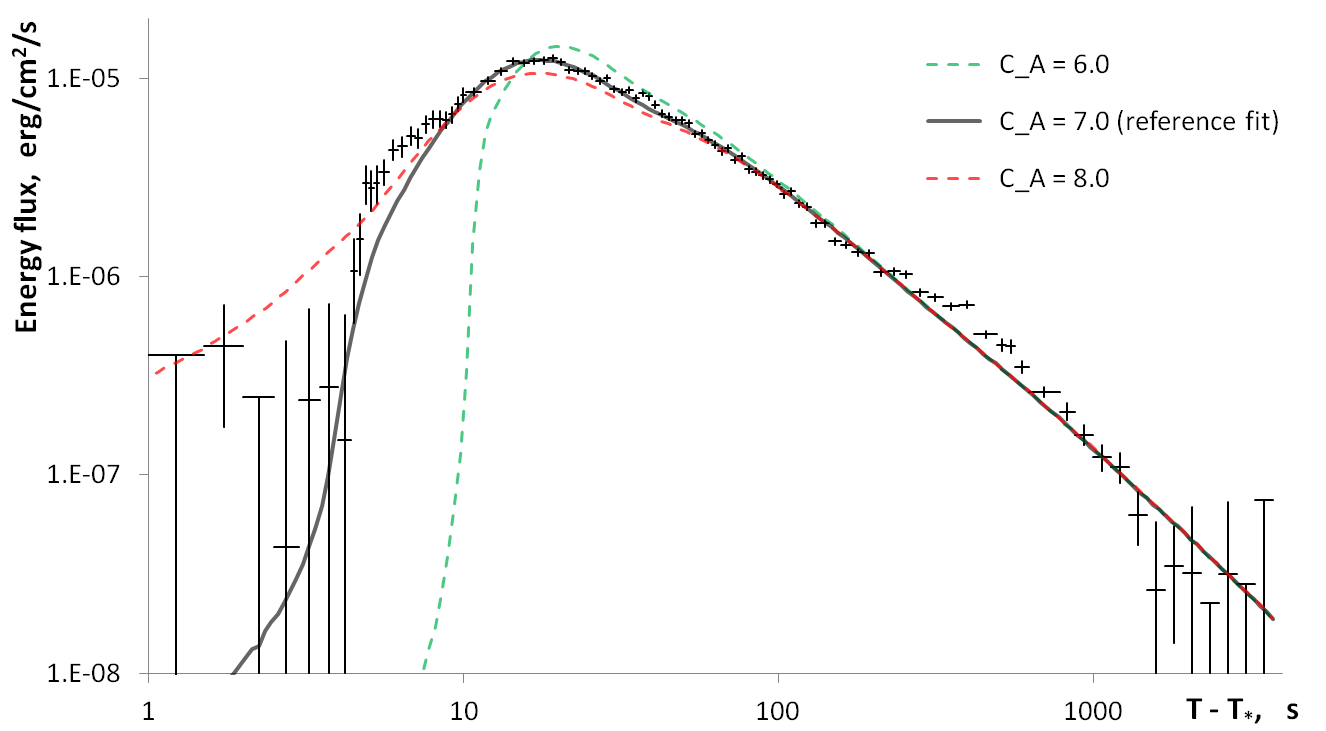}
    \includegraphics[width=1.0\linewidth]{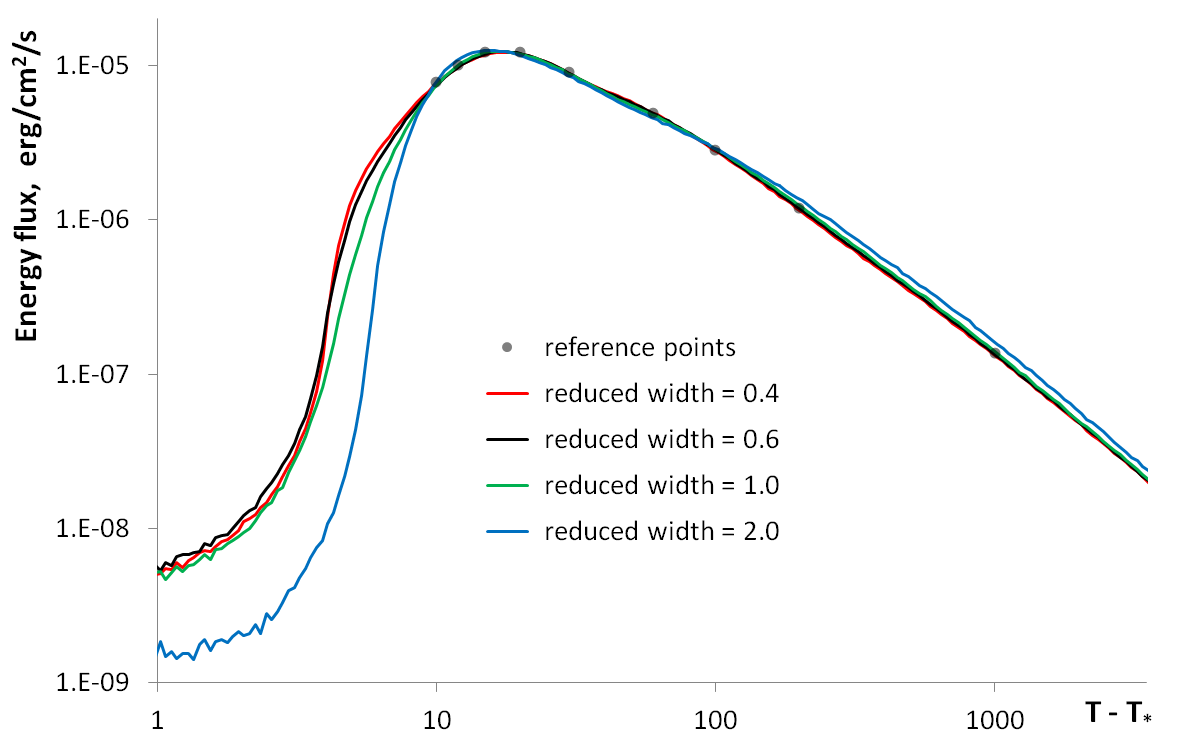}
    \caption{
    Upper panel: 
    {A comparison of our reference solution (obtained with $C_\mathrm{_A}=7.0$) to the solutions obtained with smaller ($C_\mathrm{_A}=6.0$) and larger ($C_\mathrm{_A}=8.0$) values of this coefficient. All other parameters are the same.}
    Lower panel: The reference lightcurve (black) compared to the best-fit lightcurves obtained for other values of reduced jet width $\Gamma_0 \theta_j$.
    For numerical values of parameters that characterize all the best-fit solutions see Table~(\ref{tab:fit_parameters}).
    The reference points are the same as in Fig.~(\ref{fig:TeV_lightcurve}).
    }
    \label{fig:LightcurveFits}
\end{figure}

\begin{figure}
    \centering
    \includegraphics[width=1.0\linewidth]{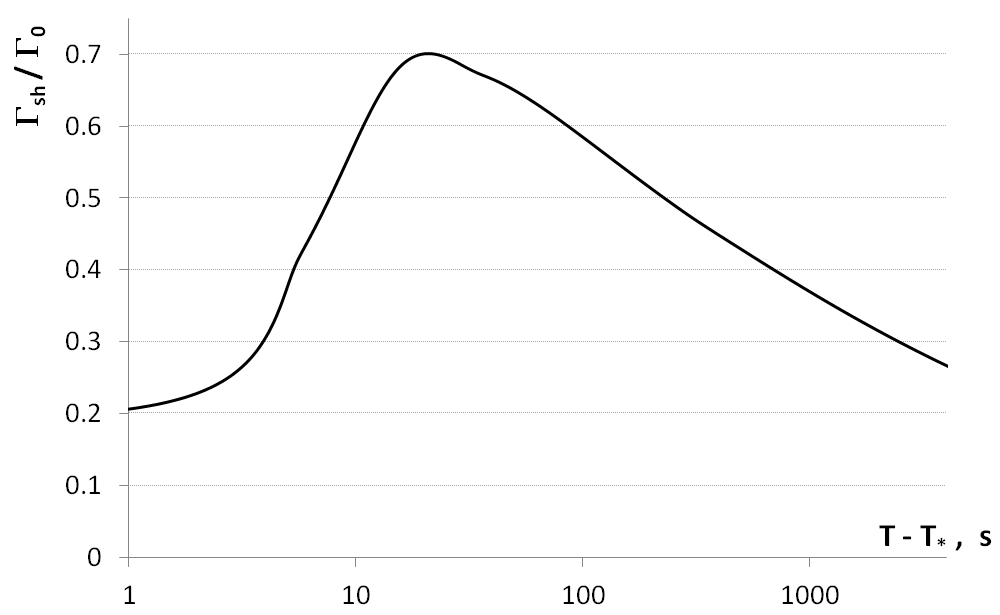}
    \caption{
    The normalized blast wave's Lorentz factor as a function of shock propagation delay (i.e., the arrival time for photons emitted straight along jet's axis),
    calculated for parameters of the reference wind-type model from Table~(\ref{tab:fit_parameters}).
    }
    \label{fig:Gamma_vs_delay}
\end{figure}

Our reference solution with 
$\Gamma_0 \theta_j = 0.6$, $C_\mathrm{_A} = 7.0$, and $t_\mathrm{d} = 130$~s
is shown in the top panel of Fig.~(\ref{fig:LightcurveFits}). 
To illustrate influence of the adjustable parameter $C_\mathrm{_A}$, we plot the reference solution together with two others, obtained with a smaller and a larger values of $C_\mathrm{_A}$. 
The bottom panel of this figure compares the reference solution with a few others. 
Figure~(\ref{fig:Gamma_vs_delay}) shows how the blast wave's Lorentz factor evolves with 
observer time (i.e., the  arrival time for photons emitted along the shock's normal)
for the reference solution. The figure shows only  a moderate decline in the Lorentz factor, about a factor 2.5    from peak to $T=3000$~s. This is consistent with the absence of spectral evolution in this temporal range.

The reference lightcurve model, as well as other good fits, has discrepancies with observational data in three regions --- at an early time ($T<9$~s), in the interval between $\approx 33$~s and $\approx 40$~s, and in the interval between $\approx 250$~s and $\approx 650$~s. In Sect.~(\ref{sec:residuals}) we argue that the discrepancies are not due to the model's deficiency but rather due to an additional signal from another source of TeV photons that closely follows the prompt activity.

Two effects mainly contribute to the puzzling fast rise of the TeV lightcurve. 
One is the rapidly increasing Lorentz boost at the time when energy associated with the largest pulse of the prompt emission is supplied to the blast wave and the latter starts to accelerate in response. Another equally important, factor is the inverse jet break phenomenon that we discussed in Sect.~(\ref{sec:model_examples}). At a very early time, when the blast wave accelerates, the observer's effective viewing angle decreases, the blast wave's emitting zone fills a larger fraction of the effective viewing angle, and the equivalent isotropic energy increases. This continues until the blast wave's Lorentz factor grows above $1/\theta_j$. Later on the emitting region fills the entire observable angle until the blast wave starts to decelerate and its Lorentz factor drops below $1/\theta_j$ thus producing the usual jet break.
For narrow jets, the inverse jet break phase continues all the way to the lightcurve's peak.

\begin{figure}
    \centering
    \includegraphics[width=1.0\linewidth]{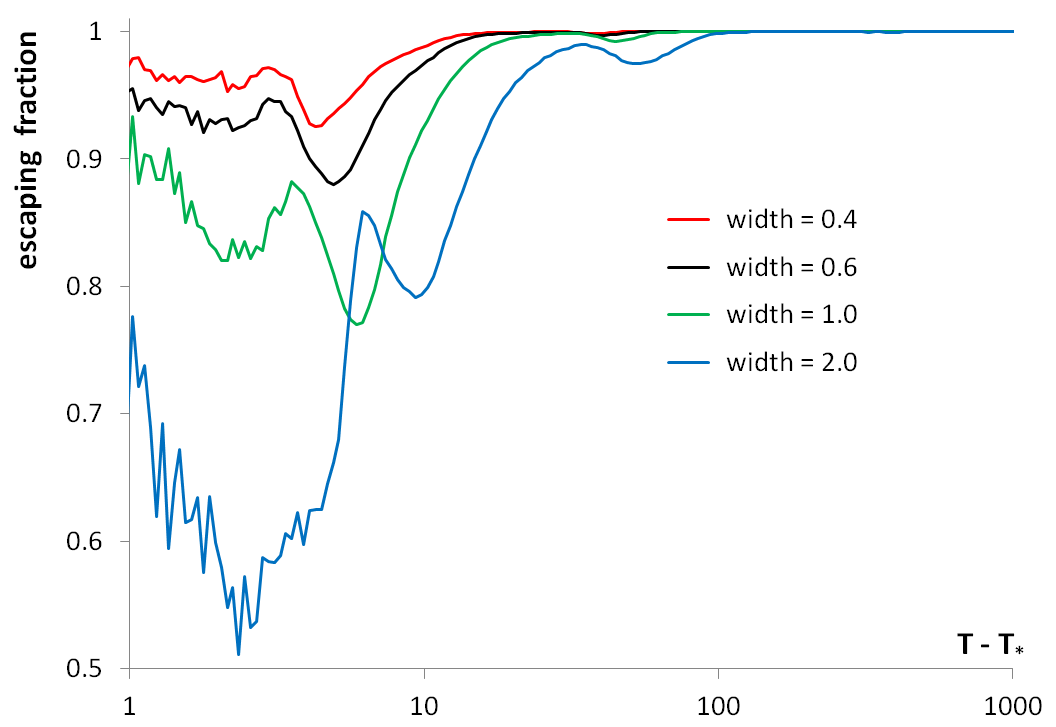}
    \caption{
    The fraction of escaping photons for best-fit solutions with different reduced width of the jet.
    In these simulations we take $\Gamma_0=500$.
    }
    \label{fig:EscapingFraction}
\end{figure}

The observed lightcurve is also
shaped by photon-photon annihilation 
 between  the outgoing afterglow photons  and those from  the prompt emission. The magnitude of this effect can be seen from Fig.~(\ref{fig:EscapingFraction}), where we plot the fraction of escaping photons (i.e. absorbed luminosity divided by unabsorbed luminosity) as a function of time. 
For the reference solution (actually for all narrow jet solutions)  the effect of 
absorption is moderate, counter to what one may naively expect. This happens for several reasons, all rooted in the fact that the target prompt photons move unidirectionally, straight along radial lines. First, the interaction rate per unit distance is proportional to $(1-\cos\theta)$, where $\theta$ is the propagation angle of the afterglow photon. The usual estimate, $\theta \sim 1/\Gamma_\mathrm{sh}$, holds for jets with not too low reduced opening angle, namely for $\Gamma_\mathrm{sh} \theta_j \gtrsim 1$. Otherwise one has to use the estimate $\theta \sim \theta_j$ instead, thus reducing the interaction rate and increasing the pair creation threshold at the same time. This effect is clearly visible when the lightcurves calculated for different jet opening angles are compared --- lightcurves for wider jets have stronger {absorption} features.
Second, the angle between the photon's momentum and the radial direction decreases as $1/R^2$ when the photon propagates outwards and, after integrating over the distance, this introduces an additional factor $1/3$ to the optical depth estimate due to reduced interaction rate alone, plus the pair creation threshold rises accordingly. 
Finally, the photons that are emitted (in the shock-comoving frame) within a small solid angle $\Omega$ around the radial direction have their optical depth lower by the factor $\Omega/(4\pi)$ compared to an average photon. Hence an optical depth $\tau_\mathrm{abs}$ results in only $(1+\tau_\mathrm{abs})^{-1}$ 
rather than exponential suppression.

\subsubsection{Constant density models}
\label{sec:ISM}

\begin{figure}
    \centering
    \includegraphics[width=1.0\linewidth]{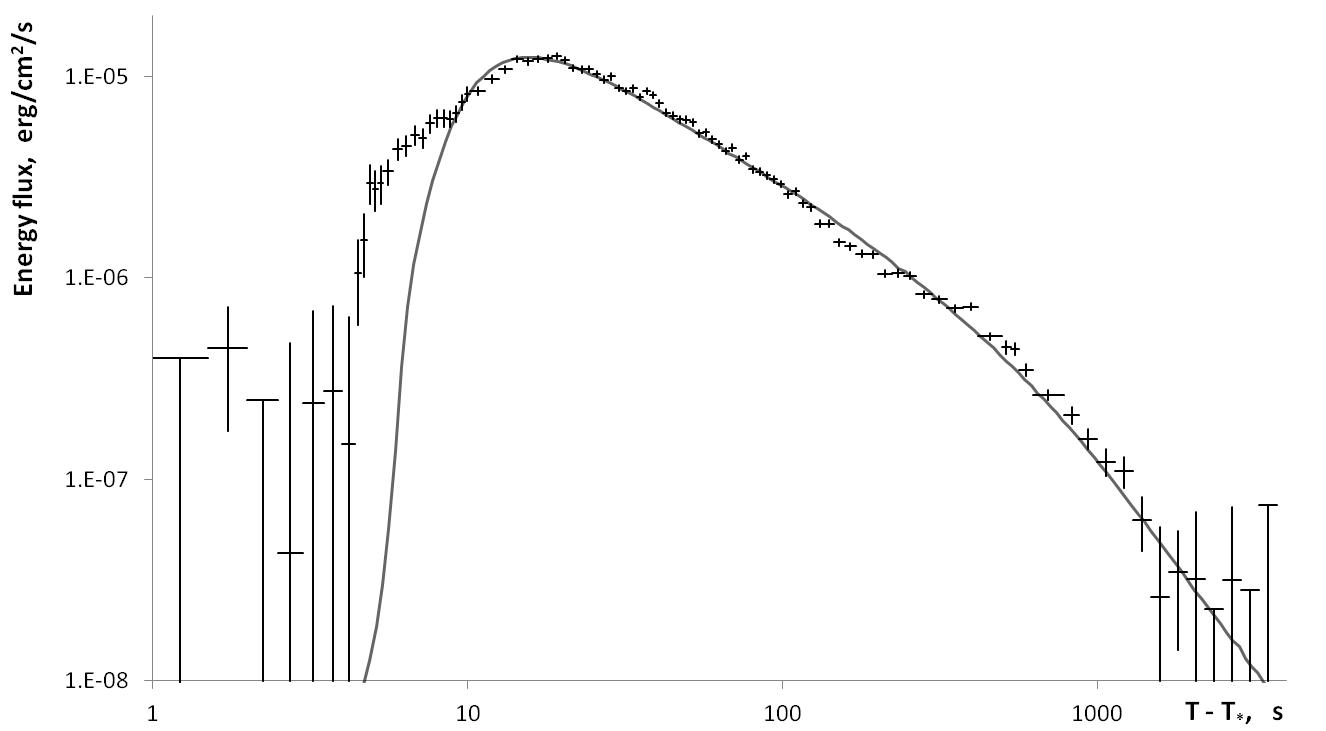}
    \caption{
    The best fit of an ISM-type lightcurve to the observational data. This solution has $t_\mathrm{d} = 4.5$~s, $\Gamma_0 \theta_j = 7$, $\Gamma_0 = 1000$, and $C_\mathrm{_A} = 14.2$.
    }
    \label{fig:ISM_fit}
\end{figure}

Models with a constant-density circumburst medium produce significantly worse lightcurve fits compared to wind-type models.
Since the peak in ISM-type solutions is determined by the blast wave's deceleration time (see Sect.~\ref{sec:model_examples}), an ISM-type model for the GRB~221009A afterglow must have small $t_\mathrm{d}$. In order to reproduce the plateau phase in the effective energy plot, clearly visible in the observational data (see bottom panel of Fig.~\ref{fig:TeV_lightcurve}) and extending to $\approx 120$~s, a jet with a large reduced opening angle is necessary, so that the jet break is sufficiently delayed.
Our search for a good fit revealed exactly this --- the best ISM-type solution has parameters 
$\Gamma_0 \theta_j = 7$, $C_\mathrm{_A} = 14.2$, $t_\mathrm{d} = 4.5$~s, and $\Gamma_0 = 1000$; it is shown in Fig.~(\ref{fig:ISM_fit}). 
The ISM-type solution is obviously a much worse fit compared to our reference wind-type model --- the residuals are much larger, especially around the peak and before it.
Moreover, these residuals don't correlate with the prompt activity and have regions of both excess and deficit. The most likely interpretation of such residuals is that constant-density models are not valid for  GRB~221009A.
Note also that by the time the TeV observational data end (at approx. 3700~s) the blast wave's Lorentz factor in the best-fit ISM-type model drops to about 7 percent of its value at the lightcurve's peak. It is unclear whether this dramatic change is consistent with the absence of spectral evolution in the TeV band.

\subsection{Residuals: possible evidence for the reverse shock emission} 
\label{sec:residuals}

\begin{figure}
    \centering
    \includegraphics[width=1.0\linewidth]{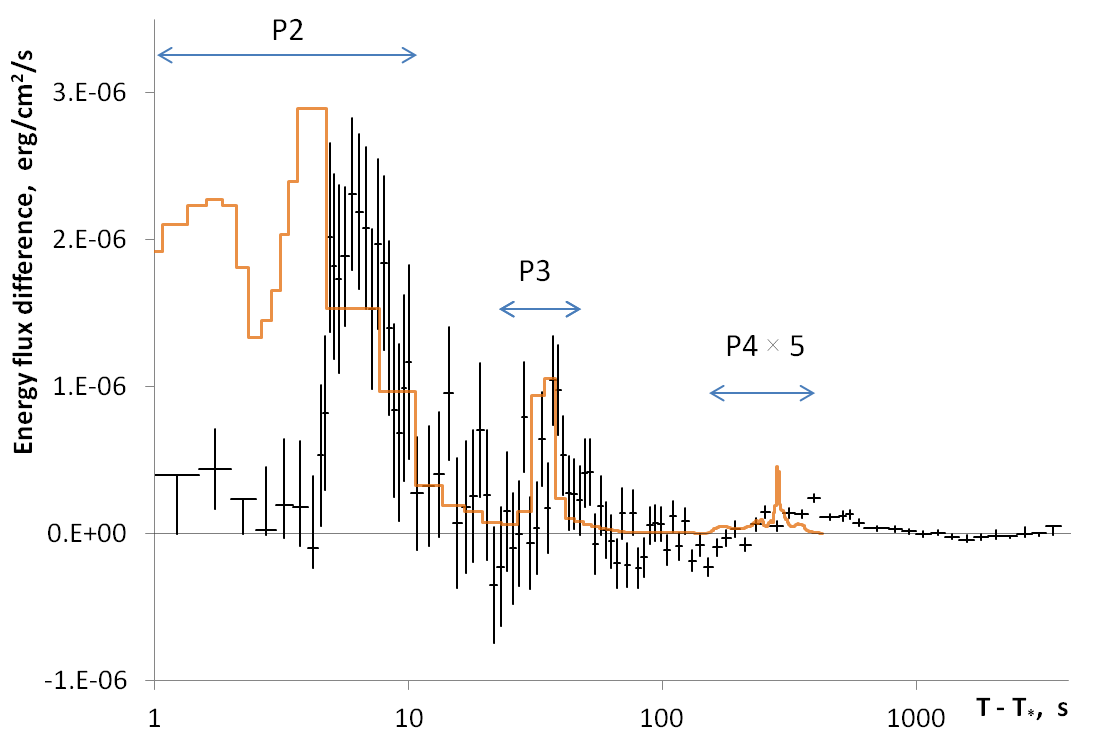}
    \caption{
    Residuals for the reference lightcurve model (black crosses) together with prompt emission count rate (orange line, arbitrary scale). 
    Note that the count rate (but not the residuals in this region) in the last pulse (P4) is not to scale -- it is multiplied by an additional factor of 5 for better visibility. When integrating the excess over time, we find that the excess associated with P4 is larger by a factor $\approx 3$ than the excesses associated with P2 and P3 combined.}
    \label{fig:Residuals}
\end{figure}

In Fig.~(\ref{fig:Residuals}) we plot the residuals for the reference lightcurve model 
$\Gamma_0 \theta_j = 0.6$, $C_\mathrm{_A} = 7.0$, and $t_\mathrm{d} = 130$~s
together with appropriately scaled count rate for the prompt phase. 
There are three regions of statistically significant excess in observational data as compared to the model. They follow with a short delay the three  largest prompt count rate pulses { (P2, P3, and P4).} 
The most natural interpretation is that the excesses are signals from the reverse shock. The brightness of the reverse shock is expected to follow the pattern of the central engine's activity (see Eq.~\ref{piston_energy_eq}), and the delay is due to the time required for the jet material to reach the reverse shock location. 
Interpreting the excesses as radiation from the reverse shock, we can {compare} its TeV
radiative {
efficiency to that of the forward shock
--- in the case of P2 and P3, the reverse shock appears to be more than an order of magnitude less efficient.
}

The lightcurve's response to the last (P4) pulse of the prompt emission is disproportionally strong. The integral excess associated with P4 exceeds the aggregated P2 and P3 excess by the factor $\approx 3$.
If this is the reverse shock signal, then the radiative efficiency of the reverse shock goes up drastically at this moment, reaching a value typical for the forward shock. 
Alternatively,  this  may be due to a larger ratio of kinetic power to the jet's prompt luminosity at P4. 
Finally, it is also possible that at this late stage 
the blast wave's structure is substantially extended
and the interaction of the additional kinetic energy of the jet with the blast wave takes a different form.

Another feature  worth mentioning  is the response to the double-peaked pulse P2. In the residuals, there is a clear signal associated with the second peak, whereas the first peak manifests itself by a much weaker signal at $T \approx 1.7$~s. 
Most likely, the intrinsic responses to the first and the second peak of P2 are comparable, but the former appears greatly attenuated due to the inverse jet break effect. At the time when the response to the first peak of P2 should appear, the blast wave  has a significantly smaller Lorentz factor and the radiation is beamed into a wider angle thus reducing the apparent brightness. This explanation implies that the jet's reduced opening angle is indeed small, $\Gamma_0 \theta_j \lesssim 1$, in agreement with the parameters of our reference solution.

While the residuals for the reference wind-type solution are correlated with the prompt activity and allow interpretation as reverse shock signal,   the much larger residuals for the ISM-type fit do not show any correlation with the prompt activity and { can}  be interpreted as a drawback of {the ISM-type solutions.}

\subsection{Inverse Compton of the prompt radiation by the external shock} 
\label{sec:ECinfluence}

TeV photons in the afterglow emission originate from Comtonization of lower-energy synchrotron photons and both components are radiated by the shock-accelerated electrons. 
If the afterglow overlaps with the prompt emission, the prompt photons are present at the place (and time) where the afterglow photons are produced. The prompt radiation is external with respect to the emission zone in the blast wave. So, the standard synchrotron-self-Compton model is to be complemented by an external Compton component. The impact of the external Compton component depends on how large is the energy density of prompt radiation in comparison to the energy density of synchrotron radiation from the  shock-accelerated electrons. 

For an estimate, consider a GRB with duration $t_\mathrm{_{GRB}}$ and an isotropic equivalent radiated energy $E^\mathrm{(iso)}_\mathrm{rad} \equiv \eta_\mathrm{rad} E^\mathrm{(iso)}_\mathrm{_{GRB}}$. In the comoving frame of the post-shock material
the energy density of the prompt radiation, calculated via average luminosity, is 
\begin{equation} \label{ext_radiation_energy_density}
    e_\mathrm{pr} = \frac{E^\mathrm{(iso)}_\mathrm{rad}}{8\pi \Gamma_\mathrm{sh}^2 R^2 c t_\mathrm{_{GRB}}} \ .
\end{equation}
It  vanishes  at large distances, where the blast wave delay with respect to light propagation time exceeds $t_\mathrm{_{GRB}}$.
The energy density of the post-shock material itself (in the comoving frame) is
\begin{equation} \label{shocked_gas_energy_density}
    e_\mathrm{sh} = 2 \Gamma_\mathrm{sh}^2 \rho c^2 \ .
\end{equation}
So, the ratio of these energy densities is
\begin{equation}  \label{ext_radiation_fraction_unsimplified}
    \epsilon_\mathrm{pr} \equiv \frac{e_\mathrm{pr}}{e_\mathrm{sh}} 
    = \frac{E^\mathrm{(iso)}_\mathrm{rad}}{16 \pi \Gamma_\mathrm{sh}^4 R^2 t_\mathrm{_{GRB}} \rho c^3}
\end{equation}

For estimation purposes, we relate the blast wave's radius $R$ to observer's time $t_\mathrm{obs}$ as $R = f_\mathrm{_R} \Gamma_\mathrm{sh}^2 c t_\mathrm{obs}$,
where the factor $f_\mathrm{_R}$ takes into account evolution of blast wave's Lorentz factor. 
\footnote{
Constant Lorentz factor results in $f_\mathrm{_R}=2$. A self-similar decelerating blast wave without energy injection results in $f_\mathrm{_R}=4$ for wind-like circumburst environment, and $f_\mathrm{_R}=8$ for a constant-density environment.
} 
Then we re-write Eq.~(\ref{ext_radiation_fraction_unsimplified}) in terms of normalized dimensionless variables ($\gamma$, $m$, and $\tau$) introduced in Sect.~(\ref{sec:model_dimensionless}):
\begin{equation} \label{ext_radiation_fraction}
    \epsilon_\mathrm{pr} 
    =  \frac{f_\mathrm{_R}  C_\mathrm{_E}}{4 (3-k)}     \frac{\eta_\mathrm{rad}}{1-\eta_\mathrm{rad}} 
    \frac{1}{\gamma^2 m}     \frac{\tau_\mathrm{obs}}{\tau_\mathrm{_{GRB}}}
    \sim      \frac{1}{\gamma^2 m}     \frac{\tau_\mathrm{obs}}{\tau_\mathrm{_{GRB}}}
\end{equation}
Here we calculate the kinetic energy of the GRB's ejecta as $E^\mathrm{(iso)}_\mathrm{kin} \equiv (1-\eta_\mathrm{rad}) E^\mathrm{(iso)}_\mathrm{_{GRB}}$. The value of $\epsilon_\mathrm{pr}$ is to be compared with the energy fraction in synchrotron radiation {produced by the afterglow } 
$\epsilon_\mathrm{sy}$, and the latter is always smaller than the energy fraction in the accelerated electrons $\epsilon_\mathrm{e}$.

The value of 
$\epsilon_\mathrm{pr}$ (given by Equation~(\ref{ext_radiation_fraction}))  depends 
on whether 
the afterglow is fast ($t_\mathrm{d} < t_\mathrm{_{GRB}}$, i.e. $\tau_\mathrm{_{GRB}} > 1$) or lagging ($\tau_\mathrm{_{GRB}} < 1$), 
and on density profile of the circumburst medium.

For a lagging afterglow in wind-like environment (this is the case of GRB~221009A) $\gamma \sim 1$ and $m \approx \tau_\mathrm{obs}$, that gives 
$\epsilon_\mathrm{pr} \sim  1/\tau_\mathrm{_{GRB}} \gtrsim 1$. 

For a fast afterglowin wind-like environment $\gamma^2 \approx 1/\tau_\mathrm{_{GRB}}$ and $m \approx \gamma^2 \tau_\mathrm{obs}$, that gives 
$\epsilon_\mathrm{pr} \sim  \tau_\mathrm{_{GRB}} \gtrsim 1$. 

For a lagging afterglow in constant-density environment $\gamma \sim 1$ and $m \approx \tau_\mathrm{obs}^3$, that gives 
$\epsilon_\mathrm{pr} \sim  1/ \left( \tau_\mathrm{_{GRB}} \tau_\mathrm{obs}^2 \right) > 1/ \tau_\mathrm{_{GRB}}^3 \gtrsim 1$. 
In this case $\epsilon_\mathrm{pr}$ is largest at the beginning of the prompt phase and lowest towards its end. It increases dramatically for very {lagging afterglows}.

For a fast afterglow in constant-density environment the blast wave decelerates for most of the time. During the deceleration phase $\gamma^2 m \approx \tau_\mathrm{obs} / \tau_\mathrm{_{GRB}}$, that gives $\epsilon_\mathrm{pr} \sim 1$. In the earliest phase, at $\tau_\mathrm{obs} < 1/\sqrt{\tau_\mathrm{_{GRB}}}$, 
additional factor $\left( \tau_\mathrm{obs}^2 \tau_\mathrm{_{GRB}} \right)^{-1}$ applies.

To sum up, in every possible situation external (i.e. prompt) radiation dominates the energy density in the afterglow's emitting zone during the earliest phase, when afterglow emission overlaps with the prompt emission. From our reference model for GRB~221009A we estimate that in this burst $\tau_\mathrm{_{GRB}} \approx 0.1$ (for the largest pulse), and therefore $\epsilon_\mathrm{pr}$ is about two orders of magnitude larger than $\epsilon_\mathrm{sy}$ (assuming fast cooling, otherwise the disproportion is even larger). Nevertheless, the observed TeV lightcurve well matches the simulated bolometric lightcurve, as if the overwhelmingly large energy density of the external radiation had no influence on Comptonization.
A possible solution to this puzzle is strong Klein-Nishina suppression of  the external Compton component.
If the  prompt radiation spectrum follows a Band function with a low-energy photon index $\alpha$ and peak energy $E_\mathrm{peak}$, then locating the afterglow synchrotron peak below 
\begin{equation} \label{noEC_condition}
    E_\mathrm{p,sy} \lesssim
    \left( \frac{\epsilon_\mathrm{sy}}{\epsilon_\mathrm{pr}} \right)^\frac{1}{\alpha+2} E_\mathrm{peak}
\end{equation}
ensures that  Klein-Nishina suppression is strong enough to make external Compton unimportant compared to self-Compton.
For the parameters of GRB~221009A ($\alpha = -0.76$ and $E_\mathrm{peak} = 3.038$~MeV, see \cite{KonusWind}) and assuming $\epsilon_\mathrm{sy} = 0.15$, Eq.~(\ref{noEC_condition}) implies  $E_\mathrm{p,sy} \lesssim 100$~keV. This limit is rather constraining, but at the same time, it is not too stringent even for the earliest afterglow.

Comptonization of $\sim 100$~keV afterglow synchrotron photons, and even more so $\sim 3$~MeV prompt photons, is indeed expected to be in the Klein-Nishina regime. Assuming $\Gamma_\mathrm{em} = 300$ for the Lorentz factor of the emitting zone, we find that electrons producing 1~TeV inverse Compton photons must have the Lorentz factor $\gamma_e \gtrsim 10^4$ in the emitting zone comoving frame. This means that possible target photons are being upscattered in the Klein-Nishina regime if their energy exceeds $1~\mathrm{TeV}/\gamma_e^2 \approx 10~\mathrm{keV}$ (observer's frame).

\section{Discussion} 
\label{sec:Discussion}

Our most important findings are the following:

$\bullet$ 
We find that a wind-type solution is strongly preferred over an ISM-type solution. 
Our best-fit solution has a narrow jet with an
opening angle of $\theta_j = 0.6 / \Gamma_0$ and a deceleration time $t_\mathrm{d} = 130$~s. 
For $\Gamma_0 = 500$ the estimated density at the deceleration radius, $R_\mathrm{d} \approx 2 $~ly, 
is $n_\mathrm{d} \approx 3\times 10^{-3}~ {\rm cm}^{-3}$.  This corresponds to $\dot{M} \approx 1.2 \times 10^{-6} M_{\sun}/\mathrm{yr}$
with $V_\mathrm{wind}= 3 \times 10^3$~km/s. 
The density at the deceleration radius is proportional to $\Gamma_0^{-8}$
and the wind's mass-loss rate estimate is proportional to $\Gamma_0^{-4}$.
These strong dependencies enable us to limit the range of the initial Lorentz factor: 
$300 \lesssim \Gamma_0 \lesssim 800$.
Here the lower limit comes from the requirement $\dot{M} < 10^{-5} M_{\sun}/\mathrm{yr}$ and the upper limit from the requirement $n_\mathrm{d} > 10^{-4}~{\rm cm}^{-3}$.

$\bullet$ We find three time intervals, where the observed lightcurve has statistically significant deviation from the theoretical one. All these deviations are excesses, {all } are well correlated in time with the three largest pulses of the prompt emission, {all} are slightly delayed with respect to {the prompt pulses}
Combination of these 
facts motivates us to interpret the residuals as additional emission component from the reverse shock. If so, then the radiative efficiency of the reverse shock is in general much lower than that of the forward shock, 
being (in the TeV band) about 30 times less.

However, the last of the excesses, which presumably results from the last episode of the central engine's activity, apparently demonstrates a much higher radiative efficiency comparable to that of the forward shock.
This may be a signature of a more efficient particle acceleration mechanism switching on when the reverse shock becomes stronger.
Another possible interpretation of the excesses is that they result from the interaction of the jet's plasma component with free neutrons in a way suggested by \citep{NeutronComponent}: the neutrons released by the jet close to its origin subsequently decay at some distance between the central engine and the reverse shock and their decay products interact with the casting jet's material. The time delay in this case is attributed to the apparent decay time of free neutrons $\approx 880\ \mathrm{s}\ /\Gamma_n$. 
Interpretation of the excesses as TeV component of the prompt emission itself is the least, rather remote, possibility: in this case it is difficult to explain the systematic time delay, and the jet's Lorentz factor needs to be very large ($\Gamma_0 > 1000$) to avoid  two-photon annihilation of the prompt TeV photons."

$\bullet$ We discovered that the jet of GRB~221009A is extremely narrow, its reduced width is $\Gamma_0 \theta_j \approx 0.6$. 
Given a Lorentz factor of $\sim 500$ this corresponds to $0.07^{\circ}$. This value is smaller by an order of magnitude than the {value} of $0.8^{\circ}$ derived by estimating a jet break in {a constant-density} surroundings by \cite{LHAASO}.
This finding is unexpected {but, besides being the best-fit solution,} it is supported by several independent arguments. 

A likely explanation for the transition from fast rising phase to a slowly rising phase in the very early afterglow is the {inverse}  
jet break that occurs when the Lorentz factor of accelerating blast wave exceeds $1/\theta_j$ which  happens close to the maximum value of the Lorentz factor.  

An alternative explanation for the rising phase as the result of photon-photon annihilation  at early time cannot explain the entire amplitude of this phase or otherwise the absorption will form a strong depression (see Fig.~\ref{fig:EscapingFraction}) at the time of the second highest pulse in the prompt emission  between  {40 and 70 s} from $T_*$, which is not observed.
Wider jets  show a stronger {photon-photon annihilation}
(as long as {$\Gamma_\mathrm{sh} \theta_j \lesssim 1$}). The absence of {a corresponding absorption}  feature effectively sets a limit on the actual jet width, {$\theta_j < 5 \times 10^{-3}$~rad (for $\Gamma_0 = 500$).}

Note also that a small reduced opening angle $\Gamma_0 \theta_j$ leads to a higher estimate for the relative efficiency of TeV emission. This  higher TeV efficiency would put the afterglow of GRB~221009A closer to values seen in other TeV afterglows.

$\bullet$ We calculated  lightcurves for several qualitatively different scenarios. The features that may appear in the lightcurves are: the jet break and transition to the phase of declining blast wave energy, both after the lightcurve peak, and the inverse jet break before the lightcurve peak. We were able to identify all these features simultaneously in the lightcurve of GRB~221009A. The best-fit model for this GRB is wind-type, and this conclusion is supported by the notion that inverse jet breaks are characteristic for wind-type solutions (with narrow jets), but not for ISM-type solutions. Another feature specific to wind-type solutions is present in GRB~221009A: the peak of its lightcurve is confined to the largest episode of energy release despite the fact that the blast wave deceleration time is significantly longer. We, therefore, conclude that the blast wave of GRB~221009A propagates into a stellar wind, and this is the first case where such a conclusion can be made directly from the analysis of observational data.

{Towards the end of LHAASO observing time, our best-fit lightcurve (which is wind-type) declines with temporal index $\approx -1.55$, that joins consistently with the X-ray afterglow decline changing from $\approx -1.5$ (at approximately the same time) to $\approx -1.67$ (at later time) \citep{Swift}.}
For the ISM-type model one needs an additional wide component in the jet to match the relatively slow decline after the jet break \citep{OConnor2023}.

$\bullet$ We advocate the use of the 
effective energy plots (luminosity of the afterglow multiplied by time since its beginning, $E_\mathrm{eff} \equiv L t$). { The} relationship between $E_\mathrm{eff}$ and $L$ resembles in a sense the relationship between $\nu F_{\nu}$ and $F_\nu$.
Almost all the lightcurve features discussed in the previous paragraph (except the inverse jet break) are much better visible in  such plots compared to the lightcurve representation.

$\bullet$ We find that a very good fit to the data assuming no spectral evolution at all. This is unexpected: as we show in Sect.~(\ref{sec:ECinfluence}), during the contemporaneous phase of the afterglow the prompt photons always dominate energy density in the afterglow's emitting region.
The only way to avoid their vigorous interference with the SSC mechanism is strong Klein-Nishina suppression. In the case of GRB~221009A, this consideration implies that 
Klein-Nishina suppression should extend to at least down to 100~keV (in the observer's frame) placing an upper limit on location of the afterglow's synchrotron peak.

\section{Conclusions} 
\label{sec:conclusion}

In this paper 
we propose a relatively simple hydrodynamical model for a relativistic blast wave with continuous energy supply at an arbitrarily varying rate. 
We treat the blast wave as a two-element structure. 
The central engine supplies energy to the inner part (shocked ejecta material) via the reverse shock.  
As the shocked ejecta material expands, its internal energy is transferred to the shocked external {matter}.
We take into account the inertia of the shocked external material so that the pressure difference across this region determines the derivative of the blast wave's Lorentz factor.

This model was tested against the observed TeV lightcurve of GRB~221009A, which provided a unique opportunity to explore the contemporaneous phase of prompt and afterglow GRB emission. 
The  excellent  quantitative agreement between the model's predictions and the observational data, suggests  that the model is valid.
This, furthermore, lends support to the values of our best fit parameters that we find that indicated that the event was powered by a very narrow jet 
$\theta_j \approx 0.07^{\circ} (500/\Gamma_0)$ and propagated {into a wind-like} external medium. These properties may explain the huge isotropic equivalent energy observed in the prompt phase. Small excesses over the model's predictions that are correlated with the prompt suggest{presence of} TeV {signal} from the reverse shock as well.


\section*{Acknowledgements}
This work was supported by the Russian Science Foundation under grant no. 21-12-00416 [ED], Advanced ERC grant  MultiJets, ISF grant 2126/22 and by the  Simons Collaboration on Extreme Electrodynamics of Compact
Sources (SCEECS)  [TP].  


\section*{Data Availability}

The data underlying this article are available in the article.


\label{lastpage}

\end{document}